\documentclass[usenatbib]{mn2e}

\usepackage{verbatim}
\usepackage{amssymb}
\usepackage{amsmath}
\usepackage{natbib}
\usepackage{threeparttable}
\usepackage{graphicx}

\title[Accretion in giant planet circumplanetary disks]{Accretion in giant planet circumplanetary disks}
\author[Sarah L. Keith and Mark Wardle]{Sarah L. Keith $^{1,2}$\thanks{E-mail: sarah.l.keith@mq.edu.au; mark.wardle@mq.edu.au} and Mark Wardle$^{1}$\\$^{1}$Department of Physics \& Astronomy and MQ Research Centre in Astronomy, Astrophysics \& Astrophotonics, Macquarie University, \\ NSW 2109, Australia\\$^{2}$Jodrell Bank Centre for Astrophysics, The University of Manchester, Alan Turing Building, Manchester, M13 9PL, United Kingdom}
\begin{document}
\date{Accepted Year Month date Day. Received Year Month Day; in original form 
Year Month Day}
\pagerange{\pageref{firstpage}--\pageref{lastpage}} \pubyear{Year}
\maketitle
\label{firstpage}\index{}

\begin{abstract}
During the final growth phase of giant planets, accretion is thought to be controlled by a surrounding circumplanetary disk. Current astrophysical accretion disk models rely on hydromagnetic turbulence or gravitoturbulence as the source of effective viscosity within the disk. However, the magnetically-coupled accreting region in these models is so limited that the disk may not support inflow at all radii, or at the required rate.
  
Here, we examine the conditions needed for self-consistent accretion, in which the disk is susceptible to accretion driven by magnetic fields or gravitational instability. We model the disk as a Shakura-Sunyaev $\alpha$ disk and calculate the level of ionisation, the strength of coupling between the field and disk using Ohmic, Hall and Ambipolar diffusevities for both an MRI and vertical field, and the strength of gravitational instability.

We find that the standard constant-$\alpha$ disk is only coupled to the field by thermal ionisation within $30\,R_J$ with strong magnetic diffusivity prohibiting accretion through the bulk of the midplane.
In light of the failure of the constant-$\alpha$ disk to produce accretion consistent with its viscosity we drop the assumption of constant-$\alpha$ and present an alternate model in which $\alpha$ varies radially according to the level magnetic turbulence or gravitoturbulence. We find that a vertical field may drive accretion across the entire disk, whereas MRI can drive accretion out to $\sim200\,R_J$, beyond which  Toomre's $Q=1$ and gravitoturbulence dominates. The disks are relatively hot  ($T\gtrsim800\,$K), and consequently massive  ($M_{\text{disk}}\sim0.5\,M_J$).

\end{abstract}

\section{Introduction}

Gas giant planets form within a protoplanetary disk surrounding a young star \citep{1985prpl.conf..981L}. Those orbiting within $\sim100\,$au of the star form through the aggregation of  a $\sim15M_{\text{Earth}}$ solid core and subsequent gas capture from the surrounding disk \citep{1996Icar..124...62P, 2009ApJ...695L..53B}.  During the initial slow accretion phase the protoplanet envelope is thermally supported and distended. However, once the envelope mass reaches the core mass gas accretion accelerates rapidly and, unable to maintain thermal equilibrium, the envelope collapses \citep{1996Icar..124...62P,  2009Icar..199..338L}.  This `run-away' gas accretion ends once the planet is massive  enough that it accretes faster than gas can be replenished into its vicinity. Infalling gas has too much angular momentum to fall directly onto the contracted planet, and so an accretion disk, the circumplanetary disk, forms around the planet \citep{1982Icar...52...14L, 2009MNRAS.397..657A}.

In contrast to the icy conditions implied by satellite systems around Solar System giant planets,  circumplanetary disks are likely  initially hot and convective \citep{1989oeps.book..723C}. Most of the protoplanet's mass is delivered during run-away accretion and so the circumplanetary disk must support a high inflow rate during this phase.  The formation of Jupiter consistent with the giant planet formation time-scale inferred from the life-time of protoplanetary disks (life-time\,$\sim3\times 10^6 $\,years; \citealp{ 2011ARA&A..49...67W}) suggests an inflow rate of $\dot{M}\sim10^{-6}M_J/$year.  Models of the accretion phase of a circumplanetary disk include  self-luminous disks \citep{1998ApJ...508..707Q, 2003A&A...411..623F, 2003ApJ...589..578N}, Shakura-Sunyaev $\alpha$ disks (\citeauthor{2002AJ....124.3404C} \citeyear{2002AJ....124.3404C}, \citeyear{2006Natur.441..834C};  \citealp{2005AA...439.1205A, 2013arXiv1306.2276T}), time-dependent disks with MRI-Gravitational instability limit cycles \citep{2011ApJ...740L...6M, 2012ApJ...749L..37L}, and hydrodynamical simulations (\citealp{1999ApJ...526.1001L}, \citealp{2002AA...385..647D}, \citealp{2003ApJ...599..548D}, \citealp{2008ApJ...685.1220M, 2009MNRAS.397..657A,2009MNRAS.393...49A,  2009MNRAS.392..514M,  2012A&A...548A.116R, 2012ApJ...747...47T, 2013ApJ...767...63S}). The evolution of the disk associated with  the contraction of the proto-planetary envelope and changes in the mode of accretion from the protoplanetary disk have also been addressed \citep{2010AJ....140.1168W}. 

The angular momentum transport mechanism is key in determining the disk structure and evolution, however little work has been done to model the disk self-consistently with the accretion mechanism. The $\alpha$-model invokes a source of viscosity (typically hydromagnetic turbulence is suggested) however there is no guarantee that the resulting disk complies with the conditions required for viscosity, hydromagnetic or otherwise.  An exception is the time-dependent gravo-magneto outbursting cycles modelled by \citet{2012ApJ...749L..37L}, however numerical simulations suggest disks rapidly evolve away from a gravitationally unstable state.

There are a variety of candidates for the accretion mechanism, including magnetic forces, gravitational instability, thermally-driven hydrodynamical instabilities, torque from spiral waves generated by satellitesimals [see \citealp{1995ARA&A..33..505P} and \citeauthor{TurnerPPVI} (in preparation) for a review], and stellar forcing \citep{2012A&A...548A.116R}.  Magnetic fields and gravitational instability are generally considered the most promising mechanisms within the protoplanetary disk.  Magnetically-driven accretion may result from hydromagnetic turbulence produced by the magnetorotational instability (MRI; \citealt{1991ApJ...376..214B, 1995ApJ...440..742H}), centrifugally driven disk winds associated with large-scale vertical fields \citep{1982MNRAS.199..883B, 1993ApJ...410..218W}, magnetic braking \citep{2004ApJ...616..266M}, or large-scale toroidal fields \citep{2000prpl.conf..589S}. MRI turbulence has been modelled extensively (e.g., \citealp{1996ApJ...457..355G, 2004ApJ...605..321S, 2007ApJ...659..729T, 2012MNRAS.420.2419F, 2012MNRAS.422.2737W, 2013ApJ...763...99P})  and simulations of MRI transport in protoplanetary disks indicate $\alpha\sim10^{-3}$, where $\alpha$ is the Shakura-Sunyaev viscosity parameter \citep{1973A&A....24..337S,2007MNRAS.376.1740K}.  Gravitational instability occurs in massive disks and may cause fragmentation or gravitoturbulence \citep{1964ApJ...139.1217T, 2001ApJ...553..174G}.

Certain conditions are required for these mechanisms to be effective. For example, magnetic processes can only act in sufficiently ionised `active' regions, where the evolution of the magnetic field is coupled to the motion of the disk. If the ionisation fraction is too low, magnetic diffusivity decouples their motion (e.g. \citealp{2007Ap&SS.311...35W}).  In protoplanetary disks, magnetic coupling is strong enough to permit MRI accretion in two regions: (i)  layers above the midplane where cosmic rays, and stellar X-rays and UV photons penetrate, and (ii) close to the star where the disk is hot and thermally ionised \citep{1996ApJ...457..355G}. Gravitational instability requires strong self-gravity such that Toomre's stability parameter $Q\lesssim 1$, and quasi-steady gravitoturbulent accretion further requires a cooling time-scale in excess of $\sim30$ orbital time-scales (\citealp{2012MNRAS.427.2022M}, \citealp{2012MNRAS.421.3286P}). 

Existing steady-state model circumplanetary disks are not massive enough for gravitational instability, and so testing for self-consistent accretion has focussed on identifying regions which are susceptible to the MRI. \citet{2011ApJ...743...53F} determined the thickness of the magnetically-uncoupled Ohmic  midplane `dead zone' of an $\alpha$ disk for the ionisation by cosmic rays. They find that the dead zone extends up at least $2.5$ scale heights (for plasma $\beta=10^4$) with the presence of grains extending this region to even greater heights. These results agree with the recent paper by \citet{2013arXiv1306.2276T} which includes ionisation from X-rays, radioactive decay, turbulent mixing, thermal ionisation as well as cosmic rays and accounting for  Ambipolar and Ohmic diffusion. They find that  $\alpha$ disks are magnetically coupled in surface layers above $\sim3$ scale heights unless the disk is dusty and is shielded from X-rays. They also consider magnetic coupling in the Jovian analogue to the Minimum Mass Solar Nebula-the Minimum Mass Jovian Nebula (MMJN; \citealp{2003Icar..163..198M}), finding that dust must be removed for magnetically coupled surface layers. They find that thermal ionisation in actively supplied disks may permit coupling within the inner $4\,R_J$ of the midplane, although \citet{2003A&A...411..623F} suggest a larger  thermally ionised region ($r\lesssim65\,R_J$). Either way, we conclude that current $\alpha$ models of circumplanetary disks are not necessarily susceptible to the magnetically driven accretion assumed at all radii, and that magnetically active surface layers may be too high above the midplane to carry the required accreting column. 

In this paper, we probe the viability of  self-consistent steady-state accretion through the circumplanetary disk midplane, with accretion driven by magnetic fields and gravitoturbulence. We model the disk as Shakura-Sunyaev $\alpha$ disk  and solve for the disk structure self-consistently with the opacity using the \citet{2009ApJ...694.1045Z} opacity-law  (\S\ref{sec:disk_structure}). In  \S\ref{sec:thermal_ionisation} we calculate the ionisation level produced by thermal ionisation, cosmic rays, and radioactive decay, and also consider the effectiveness of  turbulent mixing \citep{2006A&A...445..223I, 2007ApJ...659..729T, 2008A&A...483..815I}, and Joule heating in resistive MRI regions \citep{2005ApJ...628L.155I, 2012ApJ...760...56M}. We determine the magnetic field strength needed for accretion by an MRI or large-scale vertical field (\S\ref{sec:B_field}), and calculate Ohmic, Hall and Ambipolar diffusivities to determine the strength of magnetic coupling (\S\ref{sec:magnetic_diffusivity}). Motivated by the failure of the standard constant-$\alpha$ disk (\S\ref{sec:const_alpha_model}) to produce magnetic coupling consistent with the assumed viscosity profile we present an alternate $\alpha$ disk  (\S\ref{sec:thermally_ionised_model}) in which the level of magnetic transport (i.e., $\alpha$) varies radially consistent with the level of viscosity proceed by either magnetic forces or gravitational instability, as per the \citet{2002ApJ...577..534S} prescription for $\alpha$ for non-ideal magnetic transport. We 
present the results in \S\ref{sec:results}, with a summary and discussion of findings in \S\ref{sec:discussion}.

\section{Disk structure}
\label{sec:disk_structure}

We model a circumplanetary disk as an axisymmetric, cylindrical, radiative, thin disk surrounding a protoplanet of mass~$M$, in orbit around a star of mass~$M_*$, at an orbital distance~$d$.  The disk extends out to a radius~$r=R_H/3$ around the planet, where
\begin{eqnarray}
R_H&=&d\left(\frac{M}{3 M_*}\right)^{\frac{1}{3}}\nonumber\\
&\approx&743\,R_J \left(\frac{d}{5.2\,\text{au}}\right) \left(\frac{M}{M_J}\right)^{\frac{1}{3}}\left(\frac{M_*}{M_\odot}\right)^{-\frac{1}{3}}
\end{eqnarray}
is the Hill radius, $R_J$ is the radius of Jupiter, $M_J$ is the mass of Jupiter, and $M_\odot$ is the mass of the Sun
\citep{1998ApJ...508..707Q,2011MNRAS.413.1447M}.
 
The scale height, $H$, is determined by a balance between thermal pressure, the planet's gravity, and self-gravity of the disk. Toomre's $Q$ quantifies the strength of self-gravity, \citep{1964ApJ...139.1217T}
\begin{eqnarray}
 Q&=&\frac{c_s \Omega}{\pi G \Sigma}\nonumber\\
&\approx&5.3\times10^{3}\,\left(\frac{T}{10^3\,\text{K}}\right)^{\frac{1}{2}}\left(\frac{\Sigma}{10^2\text{g\,cm}^{-2}}\right)^{-1}\left(\frac{M}{M_J}\right)^{\frac{1}{2}}\nonumber\\
&&\times\left(\frac{r}{10^2\,R_J}\right)^{-\frac{3}{2}},
\label{eq:Q}
\end{eqnarray}
with $Q\gg1$ for negligible self-gravity and $Q\ll1$ for strong self-gravity. Here, $\Sigma$ is the column density, $\Omega$ is the Keplerian angular velocity, 
\begin{equation}
  \label{eq:keplerian}
\Omega=\sqrt{\frac{G M}{r^3}}\approx5.9\times10^{-7}\,\text{s}^{-1}\,\left(\frac{r}{10^2\,R_J}\right)^{-\frac{3}{2}}\left(\frac{M}{M_J}\right)^{\frac{1}{2}},
\end{equation}
 $c_s=\sqrt{kT/m_n}\approx1.9$\,km\,s$^{-1}\sqrt{T/1000\,\text{K}}$ is the isothermal sound speed with $m_n=2.34 m_p$ the mean neutral particle mass for a H/He gas at temperature $T$, $m_p$ the proton mass, and $k$ is Boltzmann's constant. 
Solving for the scale height for arbitrary $Q$ is complex [e.g, see \citealp{1978AcA....28...91P}], and so we adopt the simplified equation of vertical equilibrium (c.f., \citealp{2002ApJ...580..987K})
\begin{equation}
\Omega^2 H^2 + \pi G H \Sigma - c_s^2 = 0,
\label{eq:Hrelation_withselfgravity}
\end{equation}
with solution
\begin{equation}
 H=\frac{2Q}{1+\sqrt{1+4Q^2}}\frac{c_s}{\Omega}.
\label{eq:h_self_gravity}
\end{equation}
This reduces to the standard approximations
\begin{eqnarray}
\frac{H}{r} &=& \frac{c_s}{r\Omega}\nonumber\\
&\approx&0.45\,\left(\frac{T}{10^3\,\text{K}}\right)^{\frac{1}{2}}\left(\frac{r}{10^2\,R_J}\right)^{\frac{1}{2}}\left(\frac{M}{M_J}\right)^{-\frac{1}{2}}
\label{eq:H_noselfgravity}
\end{eqnarray}
 for low mass disks (i.e., $M_{\text{disk}}\ll M_J$) where self-gravity is negligible, and
\begin{eqnarray}
 \frac{H}{r}&=&\frac{c_s^2}{\pi G\Sigma r}=Q\frac{c_s}{r\Omega}\nonumber\\
&\approx&2.4\times10^{-2}\,\left(\frac{T}{10^2\,\text{K}}\right)\left(\frac{\Sigma}{10^6\,\text{g}\,\text{cm}^{-2}}\right)^{-1}\nonumber\\
&&\times\left(\frac{r}{10^2\,R_J}\right)^{-1}
\end{eqnarray}
for massive, cool, self-gravitating disks. 
From this we estimate the vertically-averaged neutral mass density
\begin{eqnarray} 
 \rho&=&\frac{\Sigma}{2H},\nonumber\\
&\approx&6.2\times10^{-9}\,\text{g cm}^{-3}\, \left(\frac{\Sigma}{10^2\,\text{g cm}^{-2}}\right)\left(\frac{T}{10^3\,\text{K}}\right)^{-\frac{1}{2}}\nonumber\\
&&\times\left(\frac{r}{10^2\,R_J}\right)\left(\frac{M}{M_J}\right)^{-\frac{1}{2}},
\label{eq:density}
\end{eqnarray}
and the associated number density, $ n=\rho/m_n\approx 2.6\times10^{15}\,\text{cm}^{-3}\,\left(\rho/10^{-8}\,\text{g\,cm}^{-3}\right)$.

The thermal structure of the disk is governed by dissipation driven by the inflow. We use the standard plane-parallel stellar atmosphere model \citep{1990ApJ...351..632H},
\begin{equation}
 \sigma T^4=\frac{3}{8} \tau \sigma T_s^{4},
\label{eq:radiative_transfer}
\end{equation} 
to calculate  the midplane temperature $T$ from the surface temperature $T_s$ and optical depth, $\tau$, from the midplane to the surface. 
Gravitational binding energy released during infall results in a surface temperature \citep{1981ARA&A..19..137P}
\begin{eqnarray}
 T_s&=& \left( \frac{3 \dot{M}\Omega^2}{8\pi \sigma} \right)^{\frac{1}{4}}\nonumber\\
&\approx&82\,\text{K}\left(\frac{\dot{M}}{10^{-6}\,M_J/\text{year}}\right)^{\frac{1}{4}}\left(\frac{M}{M_J}\right)^{\frac{1}{4}}\left(\frac{r}{10^2\,R_J}\right)^{-\frac{3}{4}},
\label{eq:surface_temp}
\end{eqnarray}
where $\dot{M}$ is the inflow rate, and $\sigma$ the Stefan-Boltzmann constant. We consider a uniform, steady, inward mass flux throughout the disk.  

Shock heating of infalling material colliding with the disk contributes additional heating, however it is negligible compared to that of the viscous dissipation [i.e., flux ratio: $F_\text{infall}/F_\text{viscous}<10^{-4}$; \citet{1981Icar...48..353C}].  Similarly, irradiation from the hot young planet [$T_J=500$\,K determined from pure contraction of the young planet; e.g. \citet{2002ARA&A..40..103H}] and the accretion hot spot [$T_{\text{hotspot}}=3300$\,K calculated using equation (3.3) in \citet{1977MNRAS.178..195P}] is also negligible with $F_\text{planet}/F_\text{viscous}<10^{-4}$ and $F_\text{hotspot}/F_\text{viscous}<10^{-2}$ determined using equation (21) from \citet{2013arXiv1306.2276T}.

Equations (\ref{eq:radiative_transfer}) and (\ref{eq:surface_temp}) are applicable in optically-thick regions of the disk (i.e., where optical depth $\tau\gg1$). This is appropriate for  the midplane, as the high column density  favours a large optical depth:
\begin{equation}
\label{eq:optical_depth}
\tau=\kappa\Sigma/2\gg1.
\end{equation}

\begin{table*}
\begin{center}
\caption{\label{table:opacity_law} 
Coefficient and indices, in each opacity regime, for the  opacity law $\kappa = \kappa_i \,\rho^{a}\, T^{b}$, as given in Table 3 of  \citet{1994APJ...427..987B} and Table 1 of \citet{2009ApJ...694.1045Z}. The resulting opacity has units of cm$^2$\,g$^{-1}$. See equation (\ref{eq:transition_temperatures})  and Fig. \ref{fig:opacity_boundaries} for the boundaries of the opacity regimes. }
\begin{threeparttable}
	\begin{tabular}{l c c r l l c c r}
		\hline
		\citet{1994APJ...427..987B}&&&&&\citet{2009ApJ...694.1045Z}&&&\\
		\hline
		Opacity Regime & $\kappa_i$ & $a$ & $b$ &&Opacity Regime & $\kappa_i$ & $a$ & $b$  \\
		\hline
		Ice grains & $2\times 10^{-4}$ & 0 & 2 &&Grains & $5.3\times10^{-2}$&$0$&$0.74$\\
		Evaporation of ice grains & $2\times10^{16}$ & 0 & $-7$&&Grain evaporation& $1.0\times10^{145}$&$1.3$&$-42$\\
		Metal grains & 0.1 & 0 & $1/2$&&Water vapour &$1.0\times10^{-15}$&$0$&$4.1$\\
		Evaporation of metal grains & $2 \times 10^{81}$ & 1 & $-24$ &&&$1.1\times10^{64}$&$0.68$&$-18$\\
		Molecules & $10^{-8}$ & $2/3 $ & 3 && Molecules&$5.1\times10^{-11}$&$0.50$&$3.4$\\
		H scattering & $10^{-36}$ & $1/3$ & 10&& H scattering & $8.9\times10^{-39}$&$0.38$&$11$\\
		Bound--free and free--free & $1.5\times10^{20}$ & 1 & $-5/2$&&Bound--free and free--free &$1.1\times10^{19}$&$0.93$&$-2.4$\\
		Electron scattering & 0.348 & 0 & 0 &&Electron scattering &$0.33$&$0$&$0$\\
		&&&&&Molecules and H scattering\tnote{a}&1.4&0&3.6\\
		\hline
	\end{tabular}
	\begin{tablenotes}
		\item[a] This regime is given in the footnote of Table 1 in \citet{2009ApJ...694.1045Z}.   The dominant sources of opacity  in this regime are molecular lines and H scattering (Z. Zhu 2013, private communication).	\end{tablenotes}
\end{threeparttable}
\end{center}
\end{table*}

\begin{figure}
\begin{center}
\begin{tabular}{cc}
\includegraphics[width=0.48\textwidth]{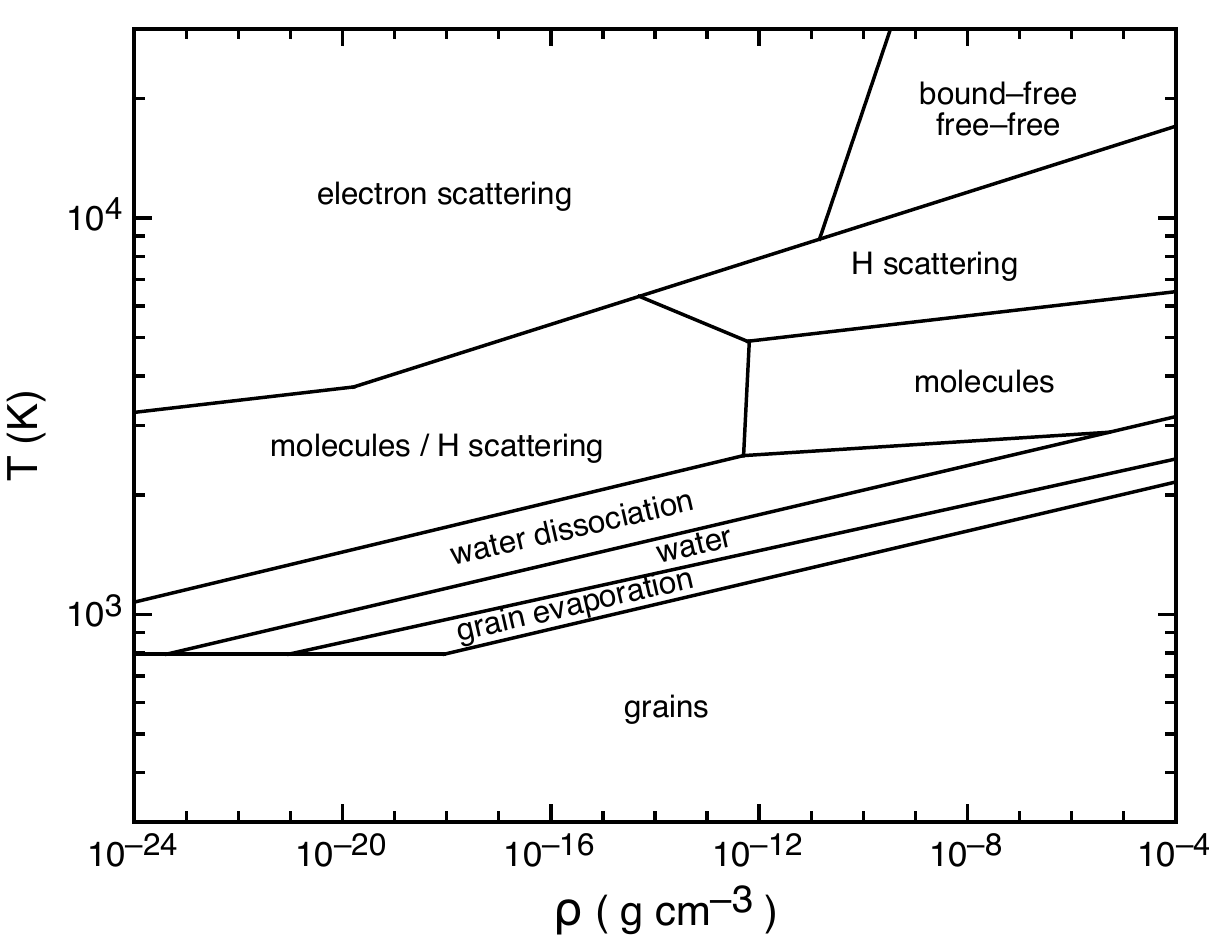}  
\end{tabular}
\end{center}
\caption{\label{fig:opacity_boundaries} Temperature and density boundaries of the \citet{2009ApJ...694.1045Z} opacity regimes, given in Table \ref{table:opacity_law}, calculated with equation (\ref{eq:transition_temperatures}).}
\end{figure}

To calculate the opacity, $\kappa$, we use the analytic Rosseland mean opacity law presented in \citet{2009ApJ...694.1045Z}. This is a piecewise power-law fit to the \citet{2007ApJ...669..483Z, 2008ApJ...684.1281Z} opacity law. We give this in Table \ref{table:opacity_law}, re-expressed as a function of temperature and density, using the ideal gas law\footnote{We have used the mean particle mass of molecular H/He gas in the conversion from pressure to density even though it is not strictly valid where hydrogen is ionised. Hydrogen is only ionised within the inner $5\,R_J$, at temperatures above $3000\,$K,  and we find that correcting the mean particle mass (to $\mu=1.24$) leads to at most a 15\% change in the temperature in this region.}. This model features nine opacity regimes, incorporating the effects of dust grains, molecules, atoms, ions and electrons. The transition temperature $T_{j\rightarrow k}$ between regimes $j$ and $k$, as a function of density, is obtained by equating the opacity in neighbouring regimes (i.e., $\kappa_j=\kappa_{k}$), and is 
\begin{equation}
T_{j\rightarrow k}=\left(\frac{\kappa_{i,j}}{\kappa_{i,k}}\right)^{\frac{1}{b_{k}-{b_j}}}\rho^{\frac{a_j-a_{k}}{b_{k}-b_j}}
\label{eq:transition_temperatures}
\end{equation}
with two additional constraints:
\begin{enumerate}
\item use Grains opacity for $T<794\,K$, and 
\item use Molecules and H scattering opacity  for $2.34\times10^4\kappa^{0.279}\,K<T<10^4\,K$.
\end{enumerate}
As the opacity law is complex we show  the temperature and density boundaries for each opacity regime in Fig. \ref{fig:opacity_boundaries}.

For comparison, we also give the \citet{1994APJ...427..987B} opacity law in Table \ref{table:opacity_law}.  This opacity law  underestimates the opacity for temperatures $T\sim$1500--3000\,K because it neglects contributions from TiO and water lines longward of 5$\mu$m \citep{1994ApJ...437..879A, 2003A&A...410..611S, 2009ApJ...694.1045Z}. The discrepancy is greatest at $\sim1700$\,K where the \citeauthor{1994APJ...427..987B} opacity is a factor $\sim500$
too low, as compared with the \citeauthor{2009ApJ...694.1045Z} model. 

We solve for the local structure (i.e., $\Sigma$ and $T$) simultaneously with the opacity, at each radius. Following \citet{1997ApJ...486..372B}, we solve for the radial temperature profile by combining equations (\ref{eq:Q}), (\ref{eq:keplerian}), (\ref{eq:h_self_gravity}), (\ref{eq:density})\,--\,(\ref{eq:optical_depth}) and the opacity law in Table \ref{table:opacity_law}, to give 
\begin{equation}
 T^{4-b}=\frac{9 \dot{M}\kappa_i}{2^{a+7}\pi \sigma}   \Omega^{2} H^{-a}\Sigma^{a+1}, 
\label{eq:temperature_density_relation}
\end{equation}
with $a$, $b$, and $\kappa_i$ specified for each opacity regime. 
This relationship allows us to describe the disk temperature and column density self consistently, when one or the other is specified.

At a given radius, we solve this equation within each opacity regime, and determine whether the resulting temperature and density fall within the limits of that regime. Solutions which do not fall within these limits are discarded. The solution is not necessarily unique, as the disk may satisfy the conditions of multiple opacity regimes (e.g., \citealp{1994APJ...427..987B, 2007ApJ...669..483Z}). 

Conservation of angular momentum provides the closing relation by specifying the accreting column needed to drive the inflow caused by turbulence, $\dot{M}=2\pi \nu\Sigma$ \footnote{ is released in a boundary layer (thickness $\ll R_J$) above the planet surface where the disk angular velocity profile transitions sharply between keplerian and the planetary rotation rate \citep{1977MNRAS.178..195P}. This contributes an additional factor $\left(1-\sqrt{R_J/r}\right)$ to the right hand side to this viscosity-inflow relation. However, we find that this factor is only significant within $r<2\,R_J$, i.e., within the boundary layer.} \citep{1973A&A....24..337S}. 
A common approach to modelling the turbulent viscosity $\nu$ is to adopt the $\alpha$-viscosity prescription, in which uncertainties in the form of the viscosity  are gathered into a single parameter $\alpha\lesssim1$ \citep{1973A&A....24..337S},
\begin{equation} 
\nu=\alpha c_s H.
\label{eq:alpha_prescription}
\end{equation}
 Observational estimates of $\alpha$, derived from the inferred mass accretion rates of T-Tauri stars, and the time dependent behaviour of FU Orionis outbursts, dwarf nova, and X-ray transients, indicate $\alpha\sim0.001-0.1$, while numerical magnetohydrodynamical shearing box simulations yield $\alpha\sim 0.01$--$10^{-3}$ [see \citet{2007MNRAS.376.1740K} and references therein].  
 This results in an accreting column 
 \begin{eqnarray}
 \Sigma&=&\frac{\dot{M}}{2\pi \alpha c_s H}\label{eq:mdot_alpha_relation}\\
 &\approx&1.6\times10^2 \text{g cm}^{-2} \left(\frac{\dot{M}}{10^{-6} M_J/\text{year}}\right)\left(\frac{\alpha}{10^{-3}}\right)^{-1}\nonumber\\
 &&\times\left(\frac{M}{M_J}\right)^{\frac{1}{2}}\left(\frac{T}{1000\,\text{K}}\right)^{-1}\left(\frac{r}{10^2\,R_J}\right)^{-\frac{3}{2}}
\end{eqnarray}
for negligible self-gravity. 

\section{Degree of ionisation}
\label{sec:thermal_ionisation}

In this section we calculate the level of ionisation at the midplane of the circumplanetary disk. The disk is too dense for the penetration of cosmic rays and X-rays down to the midplane, and so the  primary sources of ionisation are thermal ionisation and decaying radionuclides. We also consider  two further ionising mechanisms produced by the  action of MRI turbulence - the transport of ionisation from MRI active surface layers to the midplane by eddies, and ionisation from electric fields generated by MRI turbulence. 

\subsection{Thermal ionisation}
\label{sec:thermal_ionisation}
Ionisation leads to the production of electrons, ions  (with atomic number $j$), and charged dust grains with associated number density $n_e$, $n_{i,j}$, $n_g$, mass $m_e$, $m_{i,j}$, $m_g$, and charge $-q$, $+q$, $Z_g q$  respectively. Here, the grain mass and charge represent the mean value.  

 From this we define the total ion number density $n_i\equiv\sum_j n_{i,j}$, and average ion mass $m_i\equiv \left(n_i^{-1}\sum_j n_{i,j} m_{i,j}^{-1/2}\right)^{-2}$, where the summation runs over each ion species. 

To calculate the level of thermal ionisation we use the Saha equation
\begin{equation}
 \label{eq:saha}
\frac{n_e n_{i,j}}{n_{j}}= g_e \left(\frac{2\pi m_e k T}{h^3}\right)^{\frac{3}{2}} \exp\left(-\frac{\chi_j}{k T}\right), 
\end{equation}
where $n_{j}$ is the number density of neutrals with atomic number $j$,  $\chi_j$ is the ionisation potential of the $j^{\text{th}}$ ion species, $g_e=2$ is the statistical weight of an electron, and $h$ is Planck's constant. Table \ref{table:ion_species} gives the atomic weight and first ionisation energy of five key contributing elements: hydrogen, helium, sodium, magnesium, and potassium \citep{CRC}.

The exponential factor in the Saha equation gives rise to switch on/off behaviour in thermal ionisation, such that the bulk of atoms are ionised in a narrow temperature band around their ionisation temperature. Potassium has the lowest ionisation energy and is first to be ionised with an ionisation temperature of $T\sim10^3$\,K. 

\begin{table*}
\begin{center}
\caption{ \label{table:ion_species} Atomic weight, solar photospheric logarithmic abundance and abundance, and first ionisation potential for hydrogen, helium, sodium, magnesium, and potassium \citep{CRC, 2009ARA&A..47..481A}. Depletion of heavy elements by incorporation into grains is parametrised by $\delta$. }
\begin{tabular}{l c c c c c r }
\hline
Atomic  number& Element & Atomic weight& Logarithmic abundance& Abundance &Ionisation potential &Depletion \\
 & &  (amu) &   & & (eV) &(dex) \\
\hline
1 & H & 1.01  & 12.00 & 9.21$\times10^{-1}$ & 13.60 & 0 \\
2 &He &4.00&10.93&$7.84\times10^{-2}$&24.59 & 0\\
11 & Na & 22.98 & 6.24 & 1.60$\times10^{-6}$ &5.14 &$\delta$\\
12 & Mg & 24.31 & 7.60 & $3.67\times10^{-5}$ & 7.65 & $\delta$\\
19 & K  & 39.10 & 5.03 & 9.87$\times10^{-8}$ & 4.34 &$\delta$\\
\hline
\end{tabular} 
\end{center}
\end{table*}

We use solar photospheric abundances to model the elemental composition of the disk, as given in Table \ref{table:ion_species} \citep{2009ARA&A..47..481A}. However  heavy elements are encorporated into grains, reducing their gas phase abundance. We allow for depletion onto grains through a depletion factor $\delta$ (c.f., \citealp{2000ApJ...543..486S}).  The degree of depletion varies greatly between elements, however we make the simplifications that the abundance of elements other than hydrogen and helium are reduced by a constant factor, $10^{\delta}$. Grain depletion in the Orion nebula has been determined by comparing the abundances in the HII region (gas only) with that of Orion O stars (gas+dust; \citealt{1998MNRAS.295..401E}). Magnesium, a key grain constituent, is depleted at the level $\delta_{\text{Mg}}=-0.92$, which we adopt for all depleted elements. 

The abundance of the $j^{\text{th}}$ element is related to its logarithmic form, accounting for depletion onto grains: $X_j=\log_{10}(n_{j}/n_H)+12-\delta$, where the logarithmic abundance of hydrogen is defined to be $X_H=12$. 
The abundance is then $x_j=10^{X_j}/(\sum_i 10^{X_i})$, for which we take the logarithmic abundances of the remaining elements from \citet{2009ARA&A..47..481A}.

Dust grains also act to reduce the ionisation fraction by soaking up electrons, acquiring charge through the competitive sticking of electrons and ions to their surface. The net charge is found through the balance of preferential sticking of electrons due to their higher thermal velocity, with the subsequent Coulomb repulsion that develops. The average charge acquired by a dust grain is \citep{1987ApJ...320..803D}
\begin{equation}
 Z_g=\psi\tau-\frac{1}{1+\sqrt{\tau_0/\tau}}
\label{eq:grain_charge}
\end{equation} where
\begin{align}
\tau=\frac{a_g k T}{q^2},\\
\tau_0 \equiv \frac{8m_e}{\pi\mu m_p},\\
\mu\equiv \left( \frac{n_e s_e}{n_i}\right)^2\left(\frac{m_i}{m_p}\right),
\end{align}
where  $s_e$ is the electron sticking coefficient, $a_g$ the grain radius, and $\psi$ is the solution to the transcendental equation \citep{1941ApJ....93..369S}:
\begin{equation}
 1-\psi=\left(\mu \frac{m_p}{m_e}\right)^{\frac{1}{2}} e^{\psi}.
\label{eq:trancendental}
\end{equation}
We solve this using the second order approximation \citep{ArmstrongKulesza}
\begin{equation}
 \psi=1-\ln(1+y)+\frac{\ln(1+y)}{1+\ln(1+y)}\ln[(1+y^{-1}) \ln(1+y)]
\end{equation}
with $y\equiv e \sqrt{\mu m_p/m_e}$.

Charge fluctuations are small, with most grains having charge within one unit about this mean \citep{1979ApJ...232..729E}. Measurements and analytical estimates of the electron sticking coefficient suggest $s_e$ is in the range $10^{-3}$--$1$ \citep{1980PASJ...32..405U, 2010PhRvB..82l5408H}. As an approximation, we maximise the impact of grain charge removal by adopting $s_e\sim1$.

We adopt a constant gas to dust mass ratio ratio $\rho_d/\rho\equiv f_{dg}=10^{-2}$, grain size $a_g=0.1\mu$m, and grain bulk density $\rho_b=3$\,g cm$^{-3}$  \citep{1994ApJ...421..615P}. This leads to a grain number density
\begin{eqnarray}
n_g&=&\frac{ m_n f_{dg} n}{\frac{4}{3}\pi a_g^3 \rho_b}\nonumber\\
&\approx&3.1\times10^{3}\,\text{cm}^{-3}\left(\frac{n}{10^{15}\,\text{cm}^{-3}}\right)\left(\frac{f_{dg}}{10^{-2}}\right)\nonumber\\
&&\times\left(\frac{a_g}{0.1\,\mu\text{m}}\right)^{-3}\left(\frac{\rho_b}{3\,\text{g cm}^{-3}}\right)^{-1}.
\end{eqnarray}
Grain evaporation, which removes grain species, will cause spatial variation of these properties. For instance, very few grains would be present where the temperature exceeds the vaporisation temperature of iron ($T\sim1500$\,K at $\rho\sim10^{-7}\,$g\,cm$^{-3}$; \citealp{1994ApJ...421..615P}). However, we find that removing grains in this region (i.e., $f_{dg}=0$ for $r<7\,R_J$), or indeed uniformly across the disk (i.e., $f_{dg}=0$ for all $r$), has no effect on the boundary of the magnetically-coupled region owing to the overwhelming effectiveness of thermal ionisation here.

The final condition needed to determine the ionisation level is charge neutrality,
\begin{equation}
  n_i-n_e +Z_g n_g=0
\label{eq:charge}.
\end{equation}
To solve equations (\ref{eq:saha})--(\ref{eq:charge}), we use Powell's Hybrid Method for root finding \citep{Powell1970}, with the routine \texttt{fsolve} from the Python library \texttt{scipy.optimize} \citep{scipython}. This method is a modified form of Newton's Method, which checks that the residual is improved before accepting a Newton step. This optimisation allows for convergence despite the steep gradients caused by the exponential factor in the Saha equation. 

\subsection{Ionisation by decaying radionuclides, cosmic rays and X-rays}

Cosmic rays and the decay of radionuclides are the primary sources of ionisation in the outer disk where it is too cool for thermal ionisation. The short-lived radioisotope $^{26}$Al is the main contributor to ionisation by decaying radionuclides, yielding an ionisation rate $\zeta_R=7.6\times10^{-19}$\,s$^{-1}$ \citep{2009ApJ...690...69U}. Cosmic ray ionisation occurs at a rate $\zeta_{\text{CR}}=10^{-17}\,\text{s}^{-1}\exp(-\Sigma/\Sigma_{\text{CR}})$, where $\Sigma_{\text{CR}}=96\,\text{g}\,\text{cm}^{-2}$ is the attenuation depth of cosmic rays. 

X-rays from the young star will also ionise the surface layers [with $\zeta_{\text{XR}}=9.6\times10^{-17}\,\text{s}^{-1}\exp(-\Sigma/\Sigma_{\text{XR}})$ at the orbital radius of Jupiter for a star with Solar luminosity  \citep{1999ApJ...518..848I, 2008ApJ...679L.131T}],  however the X-ray attenuation depth is so small ($\Sigma_{\text{XR}}=8\,\text{g}\,\text{cm}^{-2}$) that X-rays do not reach the midplane and do not contribute to midplane ionisation or accretion [in contrast with \textit{surface} ionisation calculations by \citet{2013arXiv1306.2276T}].

Calculating the ionisation resulting from radioactive decay involves solving the coupled set of reaction rate equations for electrons, metal ions (number density $n_i$ with metal abundance $x_m$), and grains subject to charge neutrality.  Molecular ions are the first ions produced as part of the  reaction scheme, however, charge transfer to metals is so rapid  that metal ions are more abundant \citep{2011ApJ...743...53F}. We model the metals as a single species, adopting the mass, $m_i$, and abundance, $x_i$, of the most abundant metal - magnesium \citep{CRC, 2009ARA&A..47..481A}.  Free electrons and ions are formed through ionisation, and are removed through recombination (rate coefficient $k_{ei}$) and capture by grains (rate coefficients $k_{eg}$, $k_{ig}$ for electrons and ions, respectively). These processes are described by the following rate equations:
\begin{eqnarray}
    \frac{dn_i}{dt} &=& \zeta   n - k_{ei}n_{i}n_{e} 
       -k_{ig}n_{g} n_{i},    \label{eq:ion_radioactive_decay}\\
    \frac{dn_{e}}{dt} &=& \zeta   n - k_{ei}n_{i}n_{e} 
       - k_{e g }n_{g} n_{e},    \label{eq:electron_radioactive_decay}\\
    \frac{d Z_g}{dt}&=&
         k_{ig} n_{i}
        -k_{e g} n_{e}, \label{eq:grain_radioactive_decay}\\
0&=&n_i-n_e+Z_g n_g, \label{eq:neutrality_radioactive_decay}
\end{eqnarray}
for which we have neglected grain charge fluctuations  (see for example,  \citealt{1980PASJ...32..405U, 2011ApJ...743...53F}). Anticipating that the resulting ionisation fraction will be low, we make the following simplifications: (i) the average grain charge will be low and so we approximate $Z_g\approx0$ in calculating the rate coefficients $k_{ig}, k_{eg}$ and (ii) recombination is inefficient such that charge capture by grains dominates and we set $k_{ei}=0$. The charge capture rate co-efficients for neutral grains are
\begin{eqnarray}
k_{ig}&=&\pi a_g^2 \sqrt{\frac{8k_b T}{\pi m_i}}\nonumber\\
&\approx & 3.0\times10^{-5}\,\text{cm}^{3}\,\text{s}^{-1}\,\left(\frac{T}{10^3\text{\,K}}\right)^{\frac{1}{2}}\left(\frac{a_g}{0.1\mu\,\text{m}}\right)^2\nonumber\\
&&\times\left(\frac{m_i}{24.3\,m_p}\right)^{-\frac{1}{2}},\\
\label{eq:kmg}
k_{eg}&=&\pi a_g^2 \sqrt{\frac{8k_b T}{\pi m_e}}\nonumber\\
&\approx& 6.2\times10^{-3}\,\text{cm}^{3}\,\text{s}^{-1}\,\left(\frac{T}{10^3\text{\,K}}\right)^{\frac{1}{2}}\left(\frac{a_g}{0.1\mu\,\text{m}}\right)^2. 
\label{eq:keg}
\end{eqnarray}

Under these conditions the equilibrium electron and ion number density fractions are
\begin{eqnarray}
  \frac{n_e}{n}&=&\frac{\zeta }{k_{eg} n_{g}},\nonumber\\
&\approx& 5.2\times10^{-20}\,\left(\frac{T}{10^3\text{\,K}}\right)^{-\frac{1}{2}}\left(\frac{n}{10^{15}\,\text{cm}^{-3}}\right)^{-1}\nonumber\\
&&\times\left(\frac{\zeta}{10^{-18}\,\text{s}^{-1}}\right)\left(\frac{\rho_b}{3\,\text{g\,cm}^{-3}}\right)\left(\frac{f_{dg}}{10^{-2}}\right)^{-1}\nonumber\\
&&\times\left(\frac{a_g}{0.1\mu\,\text{m}}\right),\\\label{eq:ne_fraction}
\frac{n_i}{n}&=&\frac{k_{eg}}{k_{ig}}\frac{n_e}{n_n},\nonumber\\
&\approx&1.1\times10^{-17}\,\left(\frac{T}{10^3\text{\,K}}\right)^{-\frac{1}{2}}\left(\frac{n}{10^{15}\,\text{cm}^{-3}}\right)^{-1}\nonumber\\
&&\times\left(\frac{\zeta}{10^{-18}\,\text{s}^{-1}}\right)\left(\frac{\rho_b}{3\,\text{g\,cm}^{-3}}\right)\left(\frac{f_{dg}}{10^{-2}}\right)^{-1}\nonumber\\
&&\times\left(\frac{a_g}{0.1\mu\,\text{m}}\right)\left(\frac{m_i}{24.3\,m_p}\right)^{\frac{1}{2}}.\label{eq:ni_fraction}
\end{eqnarray}
We insert these values into equation (\ref{eq:neutrality_radioactive_decay}) to calculate an improved estimate of the grain charge:
\begin{eqnarray}
  Z_g&=&-\frac{n_i}{n_g}\nonumber\\
&\approx&-3.5\times 10^{-6} \,\left(\frac{T}{10^3\text{\,K}}\right)^{-\frac{1}{2}}\left(\frac{n}{10^{15}\,\text{cm}^{-3}}\right)^{-1}\nonumber\\
&&\times\left(\frac{\zeta}{10^{-18}\,\text{s}^{-1}}\right)\left(\frac{\rho_b}{3\,\text{g\,cm}^{-3}}\right)^2\left(\frac{f_{dg}}{10^{-2}}\right)^{-2}\nonumber\\
&&\times\left(\frac{a_g}{0.1\mu\,\text{m}}\right)^4\left(\frac{m_i}{24.3\,m_p}\right)^{\frac{1}{2}}.\label{eq:grain_charge_rd}
\end{eqnarray}
Charge capture by grains has removed a large fraction of the free electrons and so the average grain charge is small (validating our initial estimate, $Z_g\approx0$), and simply traces the ion density:

To calculate the charge resulting from the combined efforts of thermal ionisation, decay of radionuclides, and external ionisation sources we add the contributions linearly. A complete treatment would address the non-linear effects associated with using the combined charge particle population, rather than treating the populations as independent. However, as the drop-off of the radial thermal ionisation profile is so steep, the contribution of decaying radionuclides and cosmic rays within $r\lesssim55\,R_J$ is insignificant when compared to thermal ionisation. Similarly, thermal ionisation is highly inefficient beyond this distance, and so charge production is by radioactive decay and cosmic rays.

\subsection{Ionisation from MRI turbulence}
The action of MRI turbulence in the disk offer two further ionising mechanisms, which we describe below. We do not calculate the level of ionisation produced by these mechanisms, but rather determine their effectiveness within the circumplanetary disk. 

Eddies within MRI active surface layers caused by cosmic ray ionisation may penetrate into the underlying dead zone, transporting ionised material with them \citep{2006A&A...445..223I, 2007ApJ...659..729T, 2008A&A...483..815I}. Turbulent mixing may deliver enough ionisation into the dead zone for magnetic coupling and reactivation of the dead zone \citep{2007ApJ...659..729T}. The vertical mixing time-scale for diffusion through a scale height is \citep{2006A&A...445..223I}
\begin{equation}
\tau_D=\frac{H^2}{\nu}=(\alpha\Omega)^{-1},
\end{equation}
which is $1000$ dynamical times for Shakura-Sunyaev viscosity parameter $\alpha=10^{-3}$. However, free charges are removed through recombination and grain charge capture which lowers the ionisation fraction.  From equation (\ref{eq:electron_radioactive_decay}), we find that charges are removed on a time-scale
\begin{equation}
\tau_R=\left( k_{ei}\overline{n_i}+k_{eg}\overline{n_g}\right )^{-1},
\end{equation}
where the ion and grain number densities are vertically averaged along the path. We calculate the grain charge capture rate $k_{eg}$ for neutral grains, and the ion number density using the height averaged cosmic ray and constant radioactive decay ionisation rates assuming that ion capture by grains is small. We use a vertically uniform temperature, however we find no qualitative difference in the results using midplane or surface temperatures. For turbulent mixing to be effective in delivering ionisation to the midplane, it must be at least as rapid as charge removal (i.e., $\tau_D\gtrsim\tau_R$). Thus, we determine the effectiveness of midplane ionisation from active surface layers by comparing the charge removal and vertical mixing time-scales in \S\ref{sec:results}.

Ionisation is also produced through currents generated by the action of the MRI turbulent field \citep{2005ApJ...628L.155I}. The electric field, $E$, associated with the MRI may be able to accelerate electrons to high enough energies that they are able to ionise hydrogen in some regions. Such MRI `sustained' regions occur within the minimum mass solar nebula, reducing the vertical extent of the dead zone away from the midplane \citep{2012ApJ...760...56M}. Here we determine if self-sustained MRI occurs in circumplanetary disks. 

Joule heating is the primary mechanism for converting work done by shear [work per unit volume $W_{\text{S}}=(3/2)\alpha\Omega p$] into the electron kinetic energy. The work dissipated per unit volume by Joule heating of an equipartition current [i.e., the current $J_{\text{eq}}=cB_{\text{eq}}/(4 \pi H)$ associated with an equipartition field over a length scale $H$], is $W_J=f_{\text{fill}} f_{\text{sat}} J_{\text{eq}} E$. Here $c$ is the speed of light, $f_{\text{fill}}$ is the filling factor representing the fraction of the total volume contributing to Joule heating, and  $f_{\text{sat}}$ is the ratio of the saturation current in MRI unstable regions to the equipartition current. \citet{2012ApJ...760...56M} performed three dimensional shearing box simulations to determine the time, space, and ensemble averaged filling factor and MRI saturation current, finding $f_{\text{fill}}=0.264$  and $f_{\text{sat}}=13.1$.
The total energy available for ionisation through Joule heating is limited to the work done by shear (i.e., $W_J \le W_S$), and so the electric field strength cannot exceed \citep{2012ApJ...760...56M}\footnote{For consistency we insert our equation (\ref{eq:alpha_B_relation}) into equation (32) of \citealt{2012ApJ...760...56M}, and account for self-gravity which leads to stricter criterion, independent of plasma $\beta$: $f_{\text{whb}}=5.4\times10^{-2}$ for $Q=0$ [c.f., their equation (36)].}
\begin{equation}
E=\frac{3\alpha c_s  B_{\text{eq}}}{4 c f_{\text{fill}} f_{\text{sat}}} \left(\frac{2Q}{1+\sqrt{1+4Q^2}}\right).
\label{eq:selfsustain}
\end{equation}
 Given this restriction, we calculate the maximum electron kinetic energy, $\epsilon$, available from Joule heating \citep{2005ApJ...628L.155I},
\begin{equation}
 \epsilon=0.43 q E l \sqrt{m_n/m_e}
\end{equation}
where $l=1/(n \langle\sigma_{en}\rangle)\approx1\,\text{cm}\,(10^{15}\,\text{cm}^{-3}/n)$ is the electron mean free path, and $\langle\sigma_{en}\rangle=10^{15}\,\text{cm}^2$ is the momentum transfer rate co-efficient between elections and neutrals. For ionisation to be effective, the electron energy, $\epsilon$ must exceed the ionisation threshold of neutral particles within the disk. 

\section{Magnetic field strength}
\label{sec:B_field}
Further to a possible proto-planetary dynamo field (e.g., Jupiter's present day surface field is $4.2$\,G; \citealp{2003E&PSL.208....1S}), the disk may accrete its own field from the protoplanetary disk \citep{1998ApJ...508..707Q, 2013arXiv1306.2276T}. As both MRI and vertical fields have been modelled extensively in protoplanetary disks, we consider both field geometries in driving accretion in circumplanetary disks. We calculate the magnetic field strength, $B$, required to drive accretion at the inferred accretion rate, $\dot{M}=10^{-6} M_J/\text{year}$.
 
Three dimensional stratified and unstratified shearing box, and global MRI simulations with a net vertical flux indicate that during accretion the MRI magnetic field saturates with \citep{1995ApJ...440..742H, 2004ApJ...605..321S, 2011ApJ...730...94S, 2013ApJ...763...99P}
\begin{equation}
  \alpha\approx0.5\beta^{-1}=0.5\frac{B^2}{8\pi p},
\label{eq:alpha_B_relation}
\end{equation}
where $\beta\equiv 8\pi p/B^2$ is the plasma beta, and $p=c_s^2\rho$ is the pressure. This leads to a magnetic field strength
\begin{equation}
B_{\text{MRI}}= \sqrt{16\pi \alpha c_s^2\rho}, 
\label{eq:alpha_magnetic_field}
\end{equation}
which can be directly determined by the inflow rate as \citep{2007Ap&SS.311...35W}
\begin{eqnarray}
B_{\text{MRI}}&=&\left(\frac{\dot{M}\Omega^2}{c_s}\right)^{\frac{1}{2}}\left(\frac{1+\sqrt{1+4Q^2}}{Q}\right)\nonumber\\
&\approx&0.66\text{\,G}\left(\frac{\dot{M}}{10^{-6} M_J/\text{year}}\right)^{\frac{1}{2}}\left(\frac{M}{M_J}\right)^{1/2}\left(\frac{T}{10^3\,\text{K}}\right)^{-\frac{1}{4}}\nonumber\\
&&\times\left(\frac{r}{10^2\,R_J}\right)^{-\frac{3}{2}}\left(\frac{Q^{-1}+\sqrt{Q^{-2}+4}}{2}\right).
\label{eq:BMRI}
\end{eqnarray}

The equipartition field, $B_{\text{eq}}=\sqrt{8\pi p}$, defines the maximum field that the disk can support before magnetic pressure dominates over thermal pressure. From equation (\ref{eq:alpha_B_relation}) we see that the MRI field is sub-equipartition, satisfying 
\begin{equation}
   \frac{B_{\text{MRI}}}{B_{\text{eq}}}= \frac{v_a}{\sqrt{2}c_s}=\sqrt{2\alpha}
   \label{eq:equipartition_ratio}
\end{equation}
which is constant for a given $\alpha$, and where the Alfv\'{e}n speed is
\begin{eqnarray}
v_a&=&\frac{B}{\sqrt{4\pi \rho}},\nonumber\\ 
&\approx&8.9\times10^{-2}\,\text{km\,s}^{-1}\,\left(\frac{B}{1\text{\,G}}\right)\left(\frac{\rho}{10^{-9}\text{g\,cm}^{-3}}\right)^{-\frac{1}{2}}. 
\end{eqnarray}

Large-scale fields acting through disk winds and jets may also drive angular momentum transport and have been studied in the context of protoplanetary disks (e.g., \citealt{1993ApJ...410..218W, 1994ApJ...429..781S, 2013arXiv1301.0318B}). Magnetically-driven outflows have also been proposed for circumplanetary disks \citep{1998ApJ...508..707Q, 2003A&A...411..623F, 2006ApJ...649L.129M, 2011ApJ...730...27A}. If a vertical field drives the inflow the field strength must be at least \citep{2007Ap&SS.311...35W}
\begin{eqnarray}
B_{\text{V}}&=&\sqrt{\frac{\dot{M} \Omega}{2r}},\nonumber\\
&\approx&0.16\text{\,G}\left(\frac{\dot{M}}{10^{-6} M_J/\text{year}}\right)^{\frac{1}{2}}\left(\frac{M}{M_J}\right)^{\frac{1}{4}}\left(\frac{r}{10^2\,R_J}\right)^{-5/4}.
\end{eqnarray}

\section{Magnetic coupling}
\label{sec:magnetic_diffusivity}
We are now in a position to calculate the level of magnetic diffusivity within the disk to identify which regions of the disk are subject to magnetically-driven transport. A minimum level of interaction between the disk and the magnetic field is needed for magnetically-controlled accretion.

Collisions disrupt the gyromotion of charged species around the magnetic field. Collisions between  the electrons, ions, and neutrals occur at a rate $\nu_{ij}$  (for colliding species $i$ with $j$), with \citep{2008MNRAS.385.2269P}
\begin{eqnarray}
   \nu_{\text{ei}}&=&1.6\times10^{-2}\,\text{s}^{-1}\,\left(\frac{T}{10^3\,\text{K}}\right)^{-\frac{3}{2}}\left(\frac{n_e}{10\,\text{cm}^{-3}}\right)\nonumber \\
&&\times\left(\frac{n_n}{10^{15}\,\text{cm}^{-3}}\right),\\
\nu_{\text{en}}&=&6.7\times10^{6}\,\text{s}^{-1}\,\left(\frac{T}{10^3\,\text{K}}\right)^{-\frac{1}{2}}\left(\frac{\rho_n}{10^{-9}\,\text{g\,cm}^{-3}}\right),\nonumber\\\\
\nu_{\text{in}}&=&3.4\times10^{5}\,\text{s}^{-1}\,\left(\frac{\rho_n}{10^{-9}\,\text{g\,cm}^{-3}}\right),
\end{eqnarray}
where  $\rho_n=\rho-(\rho_i+\rho_e)$, and $n_n=\rho_n/m_n$ are the mass and number density of neutral particles, respectively.  
Electron--ions collisions are the dominant source of drag in the highly ionised inner region, however neutral drag dominates across the remainder of the disk. 
The Hall parameter for a species $j$, $\beta_j$,  quantifies the relative strength of magnetic forces and neutral drag. It is the ratio of the gyrofrequency to the neutral collision frequency \citep{2007Ap&SS.311...35W},
\begin {equation}
\beta_j=\frac{|Z_j|eB}{m_j c} \frac{1}{\nu_{jn}}.
\end {equation}
The Hall parameter is large,  $\beta_j\gg1$, when magnetic forces dominate the equation of motion, and small,  $\beta_j\ll1$, when neutral drag decouples the motion from the field. 

The Hall parameters for ions, electrons, and grains are \citep{ 1998MNRAS.298..507W, 2007Ap&SS.311...35W} 
\begin{eqnarray}
 \beta_i&\approx&4.6\times10^{-3}\left(\frac{B}{1\,G}\right)\left(\frac{n}{10^{15}\,\text{cm}^{-3}}\right)^{-1},\\
\beta_e&\approx&1.1 \left(\frac{B}{1\,G}\right)\left(\frac{n}{10^{15}\,\text{cm}^{-3}}\right)^{-1}\left(\frac{T}{10^3\,K}\right)^{-\frac{1}{2}}, \\
\beta_g&\approx&3.1\times10^{-3}\,Z_g\left(\frac{B}{1\,\text{G}}\right)\left(\frac{n}{10^{15}\,\text{cm}^{-3}}\right)^{-1}\left(\frac{T}{10^3\,K}\right)^{-\frac{1}{2}}\nonumber\\
&&\times\left(\frac{a_g}{0.1\,\mu\text{m}}\right)^{\frac{1}{2}}\left(\frac{\rho_b}{3\,\text{g\,cm}^{-3}}\right)^{\frac{1}{2}}.
\end{eqnarray}
Ions and grains, being the more massive particles, have a lower gyrofrequency, and hence a lower Hall parameter. Thus, neutral collision are more effective at decoupling ions and grains than electrons. This leads to three regimes, according to the neutral density: (a) Ohmic regime, high density: electron--ion or neutral collisions are so frequent as to decouple both  electrons and ions (i.e., $\beta_i\ll\beta_e\ll1$).  (b) Hall regime, intermediate density: neutral collisions decouple ions, but the electrons remain tied to the field (i.e., $\beta_i \ll 1 \ll \beta_e$). (c)  Ambipolar regime, low density: both the ions and electrons are coupled to the magnetic field, and drift through the neutrals. (i.e., $1\ll\beta_i\ll\beta_e$). 

In each regime collisions produce magnetic diffusivity which affects the evolution of the magnetic field through the induction equation:
\begin{eqnarray}
  \frac{\partial \bold{B}}{\partial t}&=&\nabla(\bold{v}\times \bold{B})-\nabla\times[\eta_O(\nabla\times \bold{B})+\eta_H(\nabla\times \bold{B})\times \hat{\bold{B}}]\nonumber\\
&&-\nabla\times[\eta_A(\nabla\times \bold{B})_{\perp}],
\label{eq:induction}
\end{eqnarray}
where $\bold{v}$ is the fluid velocity. The Ohmic ($\eta_O$), Hall ($\eta_H$), and Ambipolar diffusivities ($\eta_A$) are [\citealp{2008MNRAS.385.2269P}, Wardle \& Pandey (in preparation)]
\begin{eqnarray}
\label{eq:ohmic}
\eta_O&=&\frac{m_e c^2}{4\pi e^2n_e} (\nu_{\text{en}}+\nu_{\text{ei}})\nonumber\\
&\approx& 1.9\times10^{17}\text{cm}^{2}\, \text{s}^{-1}\,\left[\left(\frac{T}{10^3\,\text{K}}\right)^{\frac{1}{2}}\left(\frac{n_e}{10\,\text{cm}^{-3}}\right)^{-1} \right .\nonumber\\
&&\left .\times \left(\frac{10^{-9}\,\text{g\,cm}^{-3}}{\rho}\right)+ 2.4\times10^{-9}\,\left(\frac{T}{10^3\,\text{K}}\right)^{-\frac{3}{2}} \right ]\text{,}
\end{eqnarray}
\begin{eqnarray}
\eta_H&=&\frac{cB}{4\pi e n_e}\left(\frac{1+\beta_g^2-\beta_i^2P}{1+\left(\beta_g+\beta_i P\right)^2}\right)\nonumber\\
&\approx&5.0\times10^{17}\,\text{cm}^{2}\,\text{s}^{-1}\,\left(\frac{B}{1\,\text{G}}\right)\left(\frac{n_e}{10\,\text{cm}^{-3}}\right)^{-1}\nonumber\\
&&\times\left(\frac{1+\beta_g^2-\beta_i^2P}{1+\left(\beta_g+\beta_i P\right)^2}\right),
\label{eq:hall}
\end{eqnarray}
\begin{eqnarray}
\eta_A&=&\left(\frac{B^2}{4\pi\rho_i\nu_\text{ni}}\right)\left(\frac{\rho_n}{\rho}\right)^2\left(\frac{1+\beta_g^2+\left(1+\beta_i\beta_g\right)P}{1+\left(\beta_g+\beta_iP\right)^2}\right)\nonumber\\
&\approx&6.0\times10^{16}\,\text{cm}^{2}\,\text{s}^{-1}\left(\frac{B}{1\text{\,G}}\right)^{2}\left(\frac{\rho_n}{\rho}\right)^2\left(\frac{n_i}{10\,\text{cm}^{-3}}\right)^{-1}\nonumber\\
&&\times\left(\frac{\rho}{10^{-9}\,\text{g\,cm}^{-3}}\right)^{-1}\left(\frac{1+\beta_g^2+\left(1+\beta_i\beta_g\right)P}{1+\left(\beta_g+\beta_iP\right)^2}\right),
\label{eq:ambipolar}
\end{eqnarray}
 where $P=n_g\,\vline Z_g\vline/n_e$ is the Havnes parameter.

The magnetic field couples to the motion of the disk in regions of low magnetic diffusivity [i.e., where  $|\nabla\times(\bold{v}\times \bold{B})| \gg |\nabla\times[\eta (\nabla\times\bold{B})]|$, for each diffusivity, $\eta$].  For MRI fields we require that  the turbulent magnetic field grows faster than dissipation can destroy it such that \citep{2002ApJ...577..534S,2013ApJ...764...65M}
\begin{eqnarray}
 \eta &<& v_{a,z}^2/\Omega\nonumber\\
 &\approx&1.3\times10^{14}\,\text{cm}^{2}\,\text{s}^{-1}\,\left(\frac{B}{1\,\text{G}}\right)^2\left(\frac{\rho}{10^{-9}\text{\,g\,cm}^{-3}}\right)^{-1}\nonumber\\
&&\times\left(\frac{r}{10^2\,R_J}\right)^{\frac{3}{2}}\left(\frac{M}{M_J}\right)^{-\frac{1}{2}}
\label{eq:MRI_criterion}
\end{eqnarray}
for each diffusivity $\eta=\eta_O$, $\eta_H$, and $\eta_A$. This condition is equivalent to the condition $\Lambda>1$, where $\Lambda=v_{a,z}^2/(\eta \Omega)$ is the Elsasser number. The coupling condition uses the Alfv\'{e}n speed for the vertical component of the magnetic field. We calculate the vertical field component as $B_z\sim B_{\text{MRI}}/\sqrt{28}$, using results from \citet{2004ApJ...605..321S}.

If, instead, a vertical (rather than turbulent) field is responsible for angular momentum transport (e.g., through the action of a disk wind or jet), the condition is more relaxed as we only require that the magnetic field couples to the shear, with  \citep{2007Ap&SS.311...35W}
\begin{eqnarray}
 \eta &<& c_s^2/\Omega\nonumber\\
 &\approx &6.1\times10^{16}\,\text{cm}^{2}\,\text{s}^{-1}\,\left(\frac{T}{10^3\,\text{K}}\right)\left(\frac{r}{10^2\,R_J}\right)^{\frac{3}{2}}\left(\frac{M}{M_J}\right)^{-\frac{1}{2}}
 \label{eq:LS_criterion}
\end{eqnarray}
for each diffusivity.

Magnetic interaction still occurs for diffusivity at, or above the coupling threshold, however coupling is weak in these conditions and the connection between the dynamics of the disk and field is diminished.

\section{Disk models}
\label{sec:disk_models}
We consider four circumplanetary disk models in this paper. We present two Shakura-Sunyaev $\alpha$ disks developed for this work: (i) a constant-$\alpha$ model in which the viscosity parameter is radially uniform  (\S\ref{sec:const_alpha_model}), and (ii) a self-consistent accretion model in which the level of angular momentum transport is consistent with the strength of magnetic coupling or gravitational instability at all radii (\S\ref{sec:thermally_ionised_model}). For comparison we also describe two key circumplanetary disk models in the literature: (iii)  the Minimum Mass Jovian Nebula (\S\ref{sec:MMJN}), and (iv) the Canup \& Ward $\alpha$ disk (\S\ref{sec:canup_ward}).

\subsection{Constant-$\alpha$ model}
\label{sec:const_alpha_model}

Here we take the traditional approach, adopting the $\alpha$-viscosity prescription with a radially-uniform $\alpha$. This allows for direct comparison with existing steady state circumplanetary disk models which adopt a constant $\alpha$. We take $\alpha=10^{-3}$ in keeping with the results of simulations (with net zero magnetic flux). However, the disk may accrete a net field which enhances transport, and so we also consider $\alpha=10^{-2}$.

To obtain the radial temperature profile for this model we insert equations (\ref{eq:H_noselfgravity}) and (\ref{eq:mdot_alpha_relation}) into equation (\ref{eq:temperature_density_relation}), yielding \citep{1997ApJ...486..372B}
\begin{equation}
 T^{\frac{3}{2}a-b+5}=\frac{9\kappa_i}{2^{2a+8}\sigma }   \left(\frac{\mu m_p}{k}\right)^{\frac{3}{2}a+1} \alpha^{-(a+1)}\left(\frac{\dot{M}}{\pi}\right)^{a+2}  \left(\frac{GM}{r^3}\right)^{a+\frac{3}{2}}.
\label{eq:const_alpha_temp_density_relationship}
\end{equation}

We calculate all other properties, such as column density,  by inserting this temperature profile into the relations given in \S\ref{sec:disk_structure}.

\subsection{Self-consistent accretion model}
\label{sec:thermally_ionised_model}
The constant-$\alpha$ model implicitly assumes that the angular momentum transport mechanism operates at all radii, and to the right degree. Ionisation by cosmic rays and decaying radionuclides is insufficient to couple the disk and magnetic field \citep{2011ApJ...743...53F}, and thermal ionisation is only active in the inner disk where $T\gtrsim10^3\,$K. Without gravitoturbulence from gravitational instability, or magnetically driven transport, which relies on magnetic coupling, little if any viscosity is produced throughout the bulk of the disk  (i.e., $\alpha\approx0$). Thus, equation (\ref{eq:mdot_alpha_relation}) is invalid across the majority of the disk. 

Motivated by the inconsistency of the constant-$\alpha$ disk, we present an enhanced  steady-state $\alpha$ disk in which the level of angular momentum transport (i.e., $\alpha$) driven by magnetic fields or gravitoturbulence is consistent with the level of magnetic coupling  and strength of gravitational instability at all radii. To achieve this we divide the disk into three regions according to the mode of transport:
\begin{enumerate}
\item Saturated magnetic transport - the inner disk is hot enough for significant thermal ionisation allowing for strong magnetic coupling (i.e., $\eta_O, \eta_H, \eta_A$ are well below than the coupling threshold) and Toomre's $Q\gg1$. Magnetically-driven angular momentum transport is maximally efficient and $\alpha$ saturates at its maximum value, which we take as $\alpha_{\text{sat}}=10^{-3}$. In this region the disk is identical to the constant-$\alpha$ disk.  
\item Marginally coupled magnetic transport - in the majority of the disk, magnetic diffusivity exceeds the coupling threshold while self-gravity is still too weak for gravitoturbulence (i.e., Toomre's $Q>1$). In this intermediate region accretion is magnetically driven, although at a reduced efficiency. \citet{2002ApJ...577..534S} determined the saturation level for MRI turbulence, and hence $\alpha$, for Ohmic and Ohmic+Hall MHD simulations in the non-linear regime (i.e., $\eta\Omega/v_{a,z}^2<1$; see their Fig. 20). They find that $\alpha$ is proportional to the ratio of the coupling threshold, $v_{a,z}^2/\Omega$, to Ohmic diffusivity. By extension we also assume that the effective $\alpha$ for non-turbulent accretion (i.e., for a vertical field) also adjusts according to the level of resistivity, using the analogous coupling threshold, $c_s^2/\Omega$. Thus, in this regime for the two modes of magnetic transport, we take $\alpha$ to be
\begin{equation}
\label{eq:alphaSS02}
\alpha = \left\{
\begin{array}{lr}
\alpha_{\text{sat}} v_{a,z}^2/\left(\eta_O\Omega\right) & \text{for an MRI field,}\\
\alpha_{\text{sat}} c_s^2/\left(\eta_O\Omega\right) & \text{for a vertical field,}
\end{array}
\right.
\end{equation}
which is at most $\alpha_{\text{sat}}$ \citep{2002ApJ...577..534S}.

\item Gravoturbulent transport - in the outer disk magnetic coupling at the level required by equation (\ref{eq:alphaSS02}) would result in a gravitationally unstable disk with Toomre's $Q<1$, and so self-gravitational forces  dominate. The cooling timescale determines whether the disk fragments or enters a gravoturbulent state. We find that the cooling time-scale is much longer than the dynamical time-scale, $\Omega^{-1}$,  \citep{2007ApJ...662..642R, 2009ApJ...695L..53B} with
\begin{eqnarray}
\Omega t_{\text{cool}}&=& \frac{\Sigma c_s^2 \Omega}{\sigma T_s^4}\nonumber\\
&=&\frac{8 c_s^3}{3G \dot {M}Q}\nonumber\\
&\sim& 1.9\times10^5\, \left(\frac{T}{120\,\text{K}}\right)^{\frac{3}{2}}\left(\frac{\dot{M}}{10^{-6}\,M_J\text{/year}}\right)^{-1}Q^{-1},\label{eq:t_cool}
\end{eqnarray}
[using equations (\ref{eq:Q})  and (\ref{eq:surface_temp}), for a minimum midplane temperature $T=120\,$K  set by the temperature of the Solar Nebula  at the present day orbital radius of Jupiter according to the Minimum Mass Solar Nebula \citep{1981PThPS..70...35H}] and so gravitoturbulence rather than fragmentation occurs \citep{2012MNRAS.427.2022M}. Either by the slow build up of surface density from inflow onto the disk coupled with heating by dissipation of turbulence \citep{2001ApJ...553..174G} or by time dependent evolution of gravitationally-unstable disks \citep{2011MNRAS.410..994F, 2013ApJ...767...63S}, the disk likely evolves towards a state with $Q\sim1$. Thus, in this region we take $Q=1$.
\end{enumerate}

We solve for the disk profile by inserting equation (\ref{eq:h_self_gravity}), the scale height with self-gravity, into equation (\ref{eq:temperature_density_relation}) requiring one final relation to close the set of equations. Each region has its own closing equation to account for the differences in the mode of transport :
\begin{enumerate}
\item In the saturated magnetic transport region, we use equation (\ref{eq:mdot_alpha_relation}) with constant $\alpha=\alpha_{\text{sat}}$, inverted to give the surface density as a function of temperature.  
\item In the marginally coupled magnetic transport region we solve for the midplane temperature numerically using \texttt{fsolve} from the Python library \texttt{scipy.optimize} \citep{scipython}. The solution is determined so that $\alpha$ calculated by inverting equation (\ref{eq:mdot_alpha_relation}) is consistent with that from equation (\ref{eq:alphaSS02}). To achieve this, at each iteration of the temperature solver we calculate the surface density, scale height and Q through equations (\ref{eq:Q}), (\ref{eq:h_self_gravity}) and (\ref{eq:temperature_density_relation}) numerically using \texttt{fsolve}.  These allow us to determine $\alpha$ from equation (\ref{eq:mdot_alpha_relation}), and to also calculate the resulting ionisation fraction,  magnetic field, and diffusivity (according to \S\ref{sec:thermal_ionisation}, \S\ref{sec:B_field}, and \S\ref{sec:magnetic_diffusivity} respectively) for determining  $\alpha$  from equation (\ref{eq:alphaSS02}).  Necessarily, $\alpha$ varies radially [i.e., $\alpha\rightarrow\alpha(r)$]. 
\item In the Gravoturbulent region, we set $Q=1$ and invert equation (\ref{eq:Q}) to give the surface density as a function of temperature.  We post-calculate $\alpha(r)$ using equation (\ref{eq:mdot_alpha_relation}).
\end{enumerate}
We solve the complete set of equations using the  routine \texttt{fsolve} from the Python library \texttt{scipy.optimize} \citep{scipython}. 

\subsection{Minimum Mass Jovian Nebula}
\label{sec:MMJN}
The Minimum Mass Jovian Nebula (MMJN) is an adaptation of the Minimum Mass Solar Nebula used for modelling the Solar nebula \citep{1977ApSS..51..153W, 1981PThPS..70...35H}. The MMJN is produced by smearing out the solid mass of the satellites to form a disk, and augmenting it with enough gas to bring the composition up to solar (i.e., $f_{dg}\sim10^{-2}$). 

We use the surface density for the MMJN given in \citet{2003Icar..163..198M} which follows a $\Sigma\propto r^{-1}$ profile, except in a transition region ($20R_J<r<26R_J)$ where the profile steepens, 
\begin{displaymath}
  \Sigma=\left\{
\begin{array}{lr}
  5.1\times10^5 \text{\,g\,cm}^{-2}\left(\frac{r}{14\,R_J}\right)^{-1} &r<20\,R_J,\\
3.6\times10^{5}\text{\,g\,cm}^{-2}\left(\frac{r}{20\,R_J}\right)^{-13.5} & 20\,R_J<r<26\,R_J,\\
  3.1\times10^3 \text{\,g\,cm}^{-2}\left(\frac{r}{87\,R_J}\right)^{-1} &26\,R_J<r<150\,R_J.
\end{array}
\right.
\end{displaymath}

We use the opacity ($\kappa=10^{-4}\,\text{cm}^{2}\,\text{g}^{-1}$; appropriate for absorption by hydrogen molecules) and temperature profile given by \citet{1982Icar...52...14L},
\begin{equation}
 T = \left(240\,\text{K}\left(\frac{r}{15\,R_J}\right)^{-1}+(130\,\text{K})^{4}\right)^{1/4}.
\end{equation}
The temperature profile follows $T\propto r^{-1}$ in the optically-thick inner regions, and is matched to the temperature of the ambient nebula ($T_{\text{neb}}=130\,$K) at the outer edge of the disk. 

\subsection{Canup \& Ward $\alpha$ disk}
\label{sec:canup_ward}
Canup \& Ward (2002, 2006) model the circumplanetary disk as a steady-state, thin, axisymmetric, constant--$\alpha$ disk. They adopt the  \citet{1974MNRAS.168..603L} surface density  model, and use the  plane-parallel stellar atmosphere model to calculate the midplane temperature. Heating sources are viscous dissipation, the ambient stellar nebula ($T_{\text{neb}}=150\,$K), and  the hot young planet. The midplane temperature and density  are solved self-consistently for a uniform opacity, however a range of opacities ($\kappa=10^{-4}$--1 cm$^{2}$ g$^{-1}$) are considered to account for uncertainty in the population of sub-micron  grains. A range of inflow rates ($\dot{M}=10^{-8}$--$10^{-4} M_J/$year), and viscosity parameters ($\alpha=10^{-4}$--$10^{-2}$), are considered  to model the disk at both early and late times. However, a low inflow rate ($\dot{M}= 2\times10^{-7}M_J/$year) is needed to match the ice line with the present day location of Ganymede and to ensure solid accretion is slow enough to account for Callisto's partially differentiation. This indicates that the disk must be `gas-starved' as compared with the MMJN.
We calculate this disk model using the method given in \cite{2002AJ....124.3404C}\footnote{The profiles shown in Fig. \ref{fig:models} are calculated using the full expression $\chi=1+\frac{3}{2}[r_c/r-\frac{1}{5}]^{-1}$ (given below equation 20 in \citealt{2002AJ....124.3404C}), however we found $\chi=1$ was needed to reproduce the profiles in \citet{2002AJ....124.3404C}. For the  parameter set used here, we find that the approximation leads to at most a $37\%$ increase in the surface density, and $27\%$ reduction in the temperature profile. The difference is greatest at $r=60R_J$, but decreases toward the inner and outer boundaries.}, with parameters taken from \citet{2006Natur.441..834C} (i.e., $\alpha=6.5\times10^{-3}$,  $\dot{M}=10^{-6} M_J/$year,  and $\kappa=0.1$ cm$^{2}$\,g$^{-1}$). 

\section{Results}
\label{sec:results}
We are now in a position to apply the tools developed in \S\ref{sec:disk_structure}--\S\ref{sec:magnetic_diffusivity} to the models described in \S\ref{sec:disk_models}. All figures are shown for a protoplanet in orbit around a solar-mass star at the current orbital distance of Jupiter (i.e., $M_*=1M_{\odot}$, and $d=5.2$\,au),  calculated with the standard parameter set $\alpha=10^{-3}$, $\dot{M}=10^{-6}M_J/$year, and $M=M_J$, unless otherwise stated.

\subsection{Disk structure}

\begin{figure*}
\begin{center}
\begin{tabular}{cc}
\includegraphics[width=0.48\textwidth]{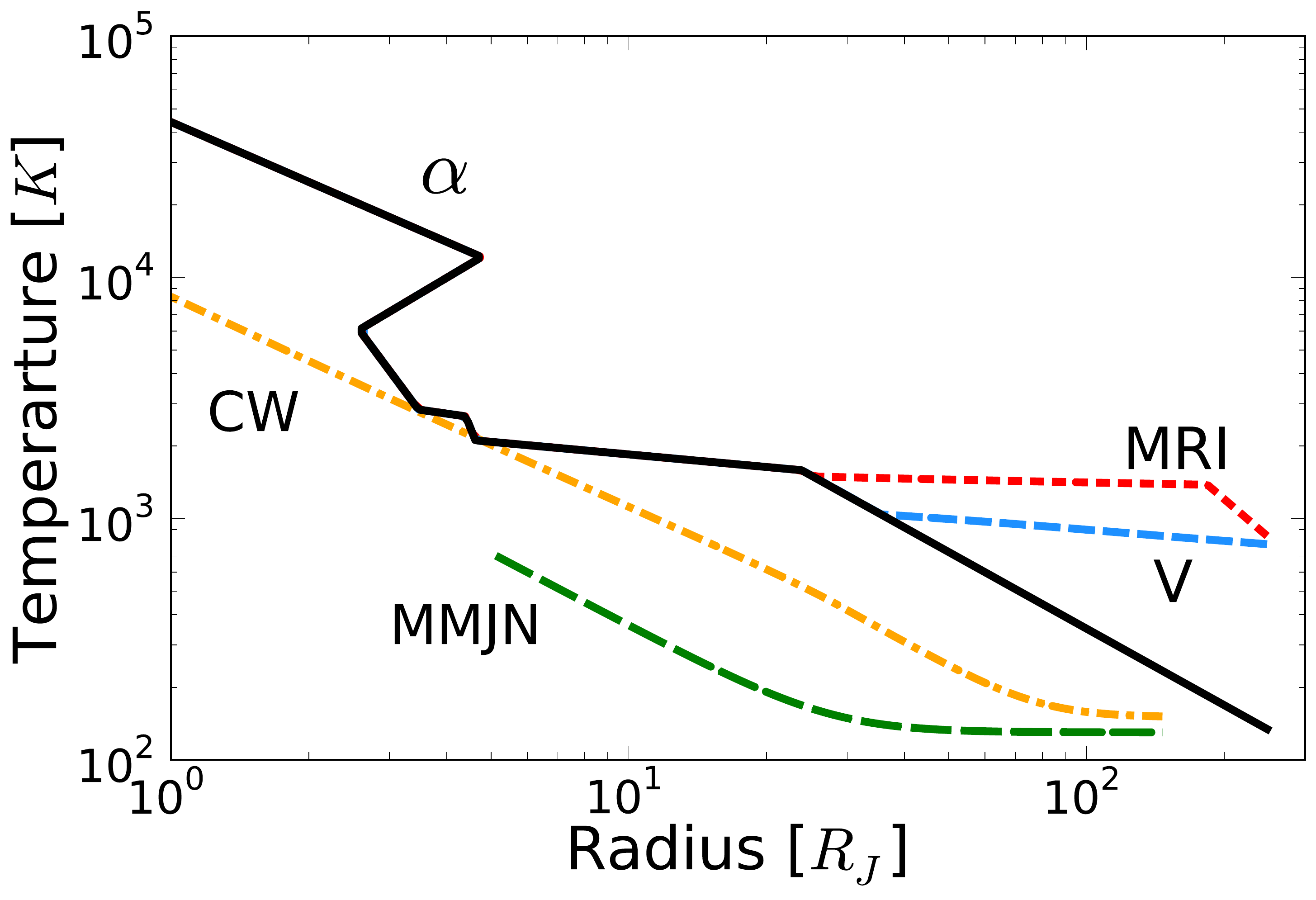}  & \includegraphics[width=0.48\textwidth]{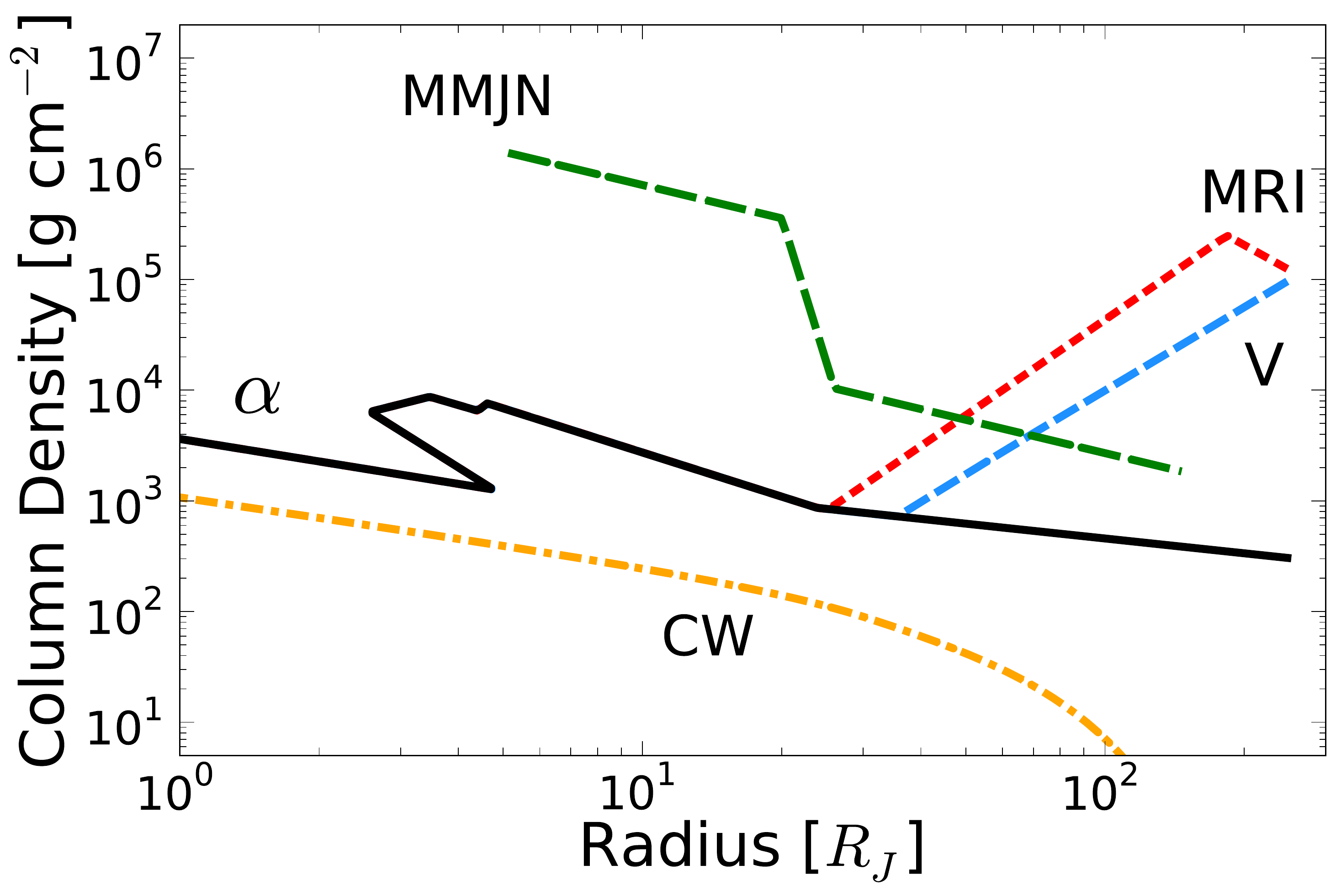} \\
\includegraphics[width=0.48\textwidth]{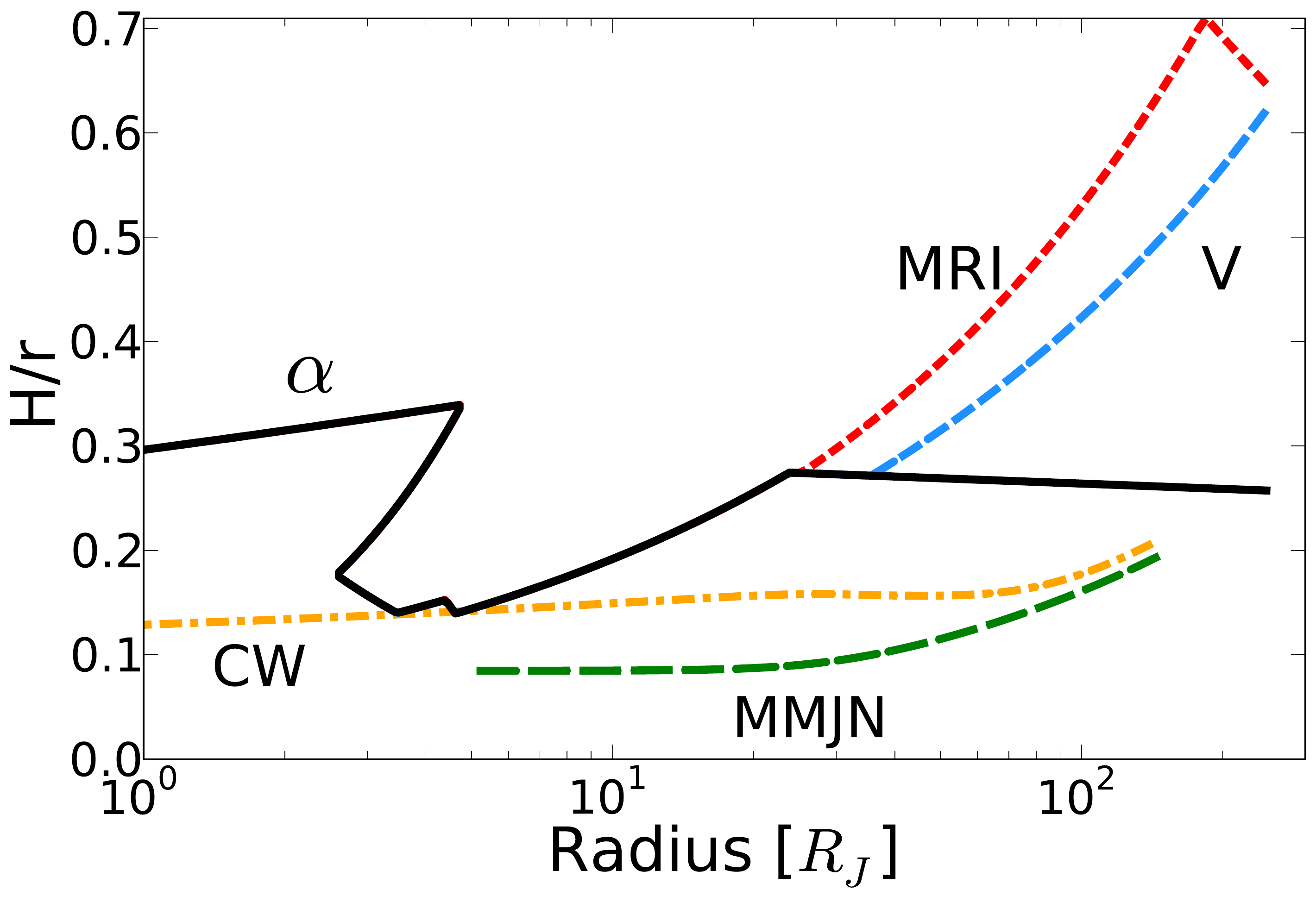}  & \includegraphics[width=0.48\textwidth]{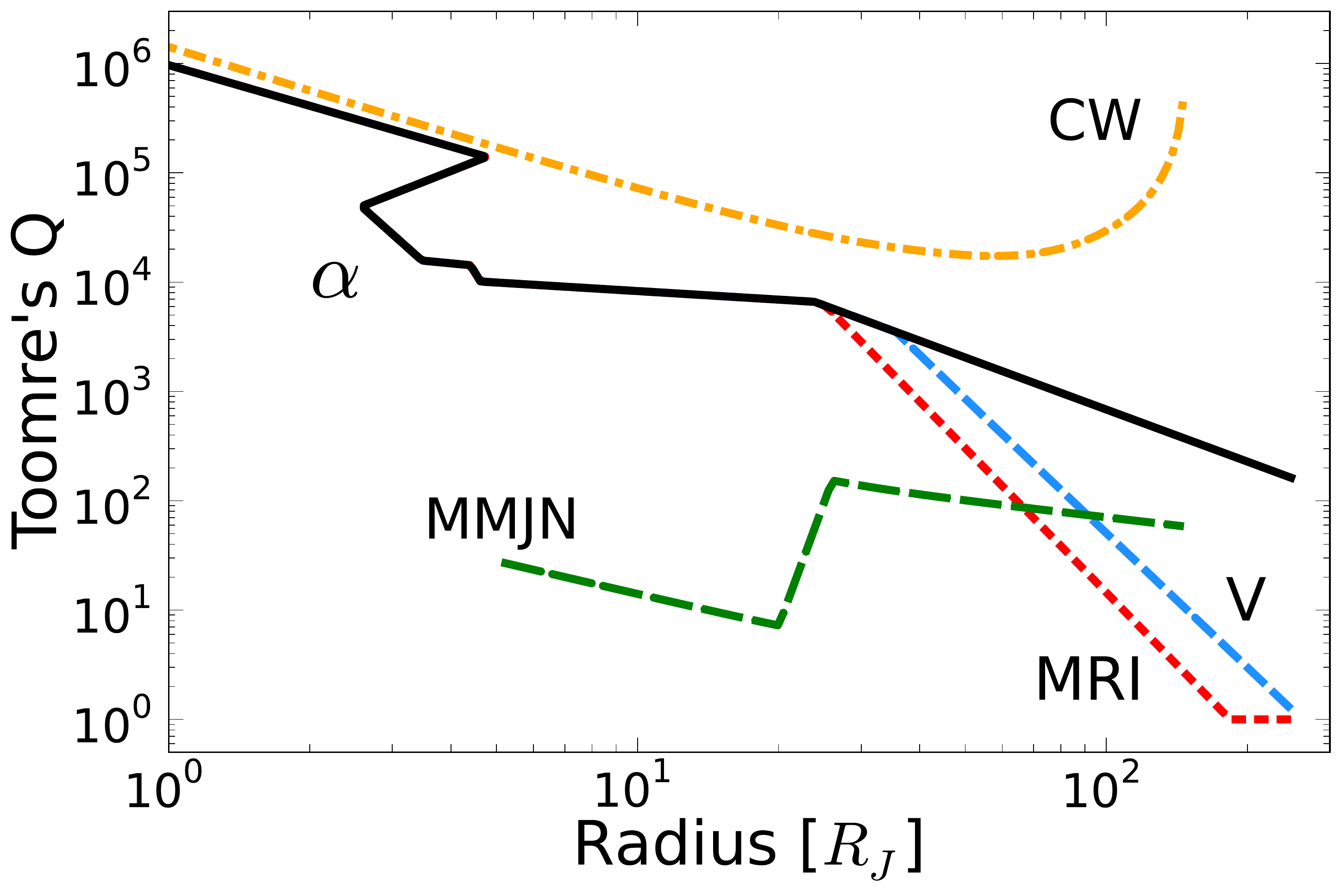}  \\
\includegraphics[width=0.48\textwidth]{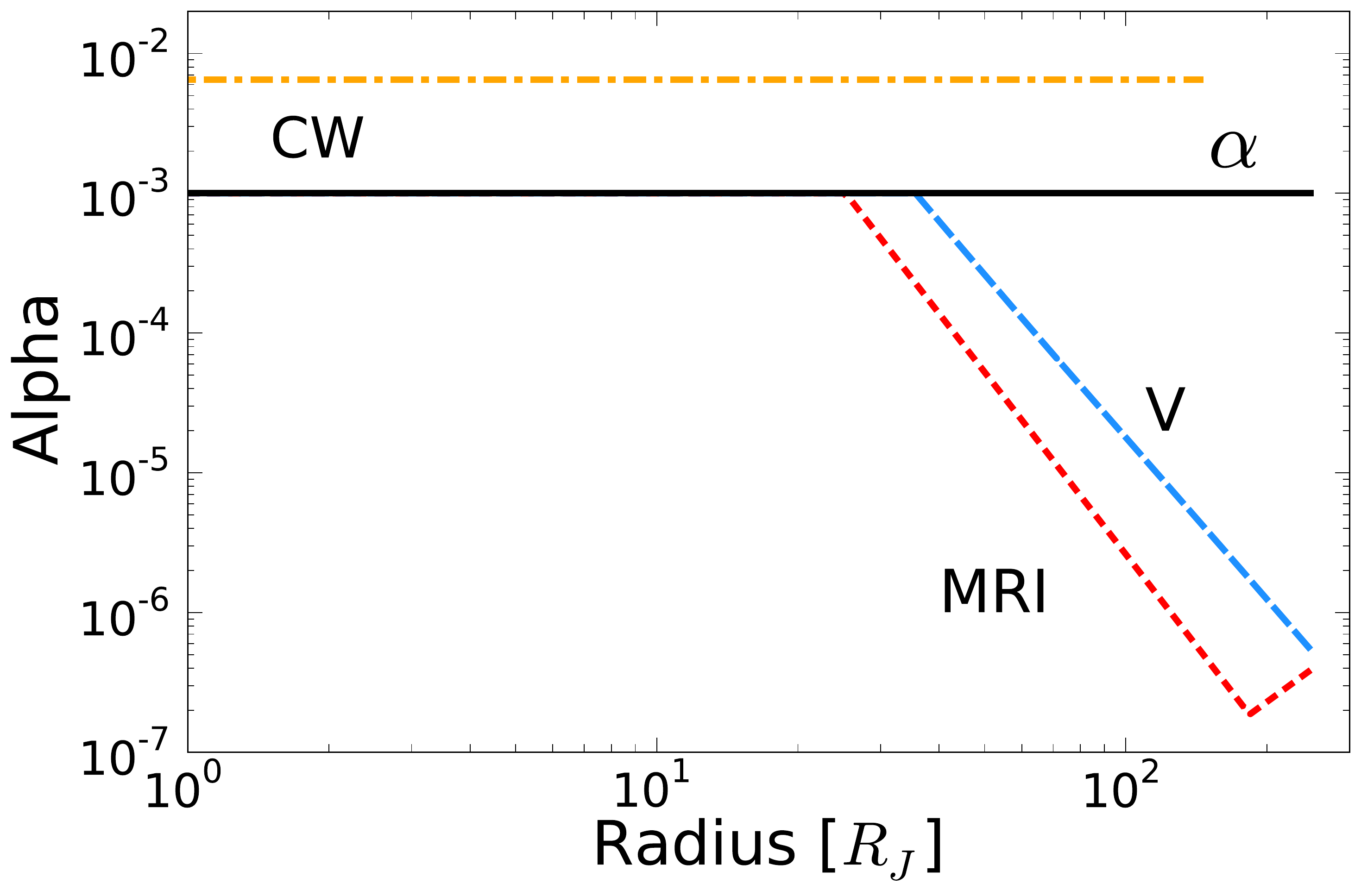}  & \includegraphics[width=0.48\textwidth]{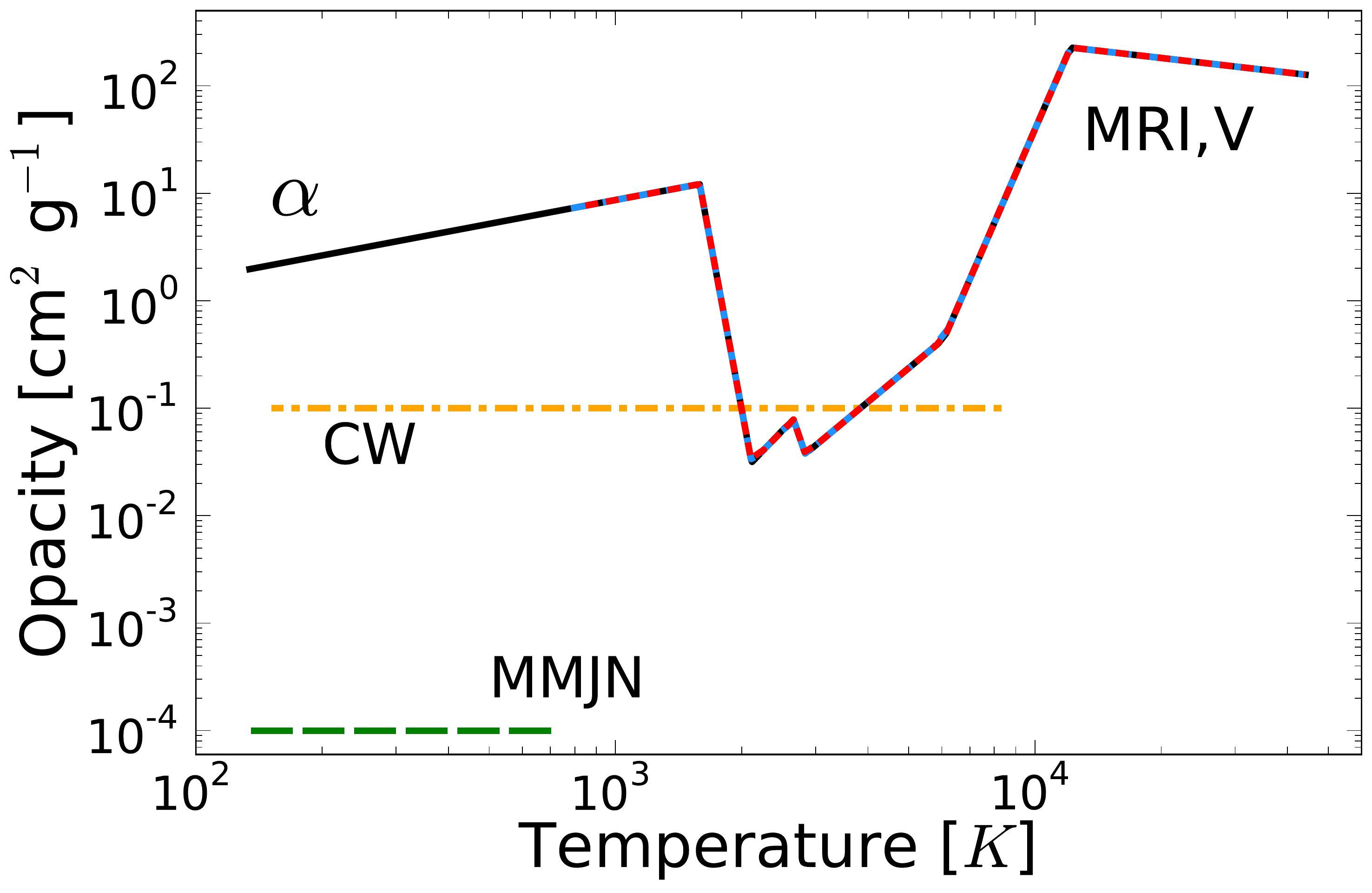}  \\
\end{tabular}
\end{center}
\caption{\label{fig:models} 
Radial dependence of the of the midplane temperature, $T$ (top-left panel), column density, $\Sigma$ (top-right panel), aspect ratio, $H/r$ (centre-left panel),  Toomre's $Q$ (centre-right panel), and viscosity parameter $\alpha$ (bottom-left panel) for  five circumplanetary disk models. Opacity as a function of temperature, $\kappa(\rho(T),T)$  is also shown (bottom-right panel).  The constant-$\alpha$ model (solid curve; \S\ref{sec:const_alpha_model}), and self-consistent accretion disk with an MRI field (dotted curve;  \S\ref{sec:thermally_ionised_model}), and vertical field (short-dashed curve; \S\ref{sec:thermally_ionised_model}) are shown  with the MMJN model (long-dashed curve; \S\ref{sec:MMJN}) and \citeauthor{2002AJ....124.3404C} disk (dot-dashed curve; \S\ref{sec:canup_ward}) for comparison.}
\end{figure*}

Fig. \ref{fig:models} shows the radial disk structure for each model. The constant-$\alpha$ disk, MMJN, and Canup \& Ward disks are shown as the solid, long-dashed, and dot-dashed curves, respectively. The self-consistent accretion disk is shown for both an MRI (dotted curve) and vertical field (short-dashed curve). The curves are labelled $\alpha$, MMJN, CW, MRI, and V respectively. 

The temperature profiles are shown in the top-left panel. The temperature profile for the constant-$\alpha$ and self-consistent accretion disks follow a power law with index changes at the transitions between opacity regimes. The self-consistent accretion disk profiles follow the constant-$\alpha$ profile out to $\sim30\,R_J$ where the temperature, and thermal ionisation level is high enough for good magnetic coupling. The stronger coupling requirement for an MRI field makes for a slightly hotter and more dense disk than for accretion driven by a vertical field, and so the disk is gravoturbulent beyond $200\,R_J$, where the temperature profile steepens. There is no corresponding gravoturbulent region for the self-consistent accretion disk with vertical field. Nevertheless, the  self-consistent accretion disk is remarkably similar when either the MRI or verticals used for drive accretion. The profile for constant-$\alpha$ disk  follows $T\propto r^{-1.1}$ in the outer regions where the opacity is primary from grains [i.e., $a=0, b=0.74$; see equation (\ref{eq:const_alpha_temp_density_relationship})]. Of the parameter set $\alpha$, $\dot{M}$ and $M$, the temperature profile is most sensitive to changes in the inflow rate. An order of magnitude change in $\dot{M}$ only corresponds to a factor $\sim3$  change in the temperature across most of the disk, with little effect beyond $\sim40\,R_J$.The profiles are multivalued in the region $r\sim2$--$5\,R_J$, with a characteristic `S-shape'. Here the disk satisfies conditions for multiple opacity regimes, with the radially increasing, unstable branch corresponding to the H-scattering opacity regime. The viscous-thermal instability associated with this feature has been used to model outbursts in circumstellar disks surrounding T-Tauri  stars - most notably FU Orionis outbursts by \citet{1997ApJ...486..372B}.

The constant-$\alpha$ and self-consistent accretion disks are hotter than the Canup \& Ward and MMJN disks, which aim to model a later phase of the disk when the opacity is from ice grains (and necessarily lower; see the opacity profile in bottom-right panel of Fig. \ref{fig:models}), and the disk is cool enough to form icy satellites. As inflow from the protoplanetary disk tapers, the disk cools, consistent with the evolution to an icy state recorded by the Solar System giant-planet satellite systems. For example, reducing the inflow rate by a factor of ten lowers the temperature to only $370\,$K at the disk outer edge. 
 
The column density profile is shown in the top-right panel. The profile for the constant-$\alpha$ disk is generally shallow, decreasing by only a factor of $\sim12$ between the inner and outer edge. Like the \citeauthor{2002AJ....124.3404C}  disk, the column density is low  compared with the MMJN, and so the disk is `gas starved'. Consequently, the disk mass is also low, with $M_{\text{disk}}=1.6\times10^{-3}M_J$, validating our neglect of self gravity.
On the other hand, the column density in the  self-consistent accretion disks increase beyond $\sim30\,R_J$  reaching $\Sigma=9.6\times10^4$\,g\,cm$^{-2}$ for a vertical field, and $\Sigma=2.5\times10^5$\,g\,cm$^{-2}$ for an MRI field. 
Consequently, the disk masses are large, with $M_{\text{disk}}=0.5\,M_J$ for the vertical field, and $M_{\text{disk}}=0.64\,M_J$ for the MRI field. The disk mass increases as the inflow rate from the protoplanetary disks tapers, such that a factor 10 reduction in the inflow rate leads to an inward extension of the gravoturbulent region, and a disk mass $M_{\text{disk}}=0.42\,M_J$, independent of the field geometry. 

The centre-left panel of Fig. \ref{fig:models} shows the aspect ratio for each model. The aspect ratio for the constant-$\alpha$ model ranges between $H/r=0.14$--$0.34$, with pressure dominating the scale height. Self-gravity  is too weak to counteract the strong thermal pressure in the  outer regions of the self-consistent accretion disks and so the disks are very thick, with the aspect ratio reaching a maximum of  $H/r=0.63$, and $0.71$ for a vertical and MRI field,  respectively. Our results agree with \cite{2013ApJ...767...63S} in that circumplanetary disks may be more aptly described as  `slim' (i.e., $H/r\lesssim1$) rather than `thin'. 

The centre-right panel of Fig. \ref{fig:models} shows the radial profile for Toomre's $Q$. Toomre's $Q$ is large for the low mass constant-$\alpha$ disk, however, despite the high temperatures the self-consistent accretion disks reach $Q\sim1$ at the outer edge where the column density is highest. We fix $Q=1$ in the gravoturbulent region in the self-consistent accretion disk with MRI field. 

The bottom-left panel of Fig. \ref{fig:models} shows the radial profile of the viscosity parameter, $\alpha$. The viscosity parameter is constant across the Canup \& Ward and constant-$\alpha$ disks, and in the inner regions of the self-consistent accretion disks where magnetic coupling is good and $\alpha$ is saturates at its maximum value. Once the temperature drops below $\sim1000\,K$ thermal ionisation drops and with it the strength of magnetic coupling. Magnetic transport is less efficient with high diffusivity and so $\alpha$ is  reduced, as per equation (\ref{eq:alphaSS02}), reaching a minimum of  $1.9\times10^{-7}$ for an MRI field, and $4.8\times10^{-7}$ for a vertical field. In the outer $\sim60\,R_J$ of the self-consistent MRI accretion disk, $\alpha$ increases radially to compensate for the decreasing column density. However, such a low  required effective viscosity is potentially overwhelmed by  other processes, such as stellar forcing  or satellitesimal wakes which may contribute additional torque exceeding  this level \citep{2012A&A...548A.116R, 2001ApJ...552..793G}. 

Note that a property of this model is that temperature increases with decreasing $\alpha$. This result is counter intuitive given that viscosity, and hence dissipation, are directly proportional to $\alpha$. However, for a fixed $\dot{M}$,  increasing $\alpha$ enhances the effectiveness of the turbulence and so reduces the required active column density [see equation (\ref{eq:mdot_alpha_relation})]. The associated reduction in optical depth lowers the midplane temperature relative to the surface temperature. Consequently, if we increase $\alpha_{\text{sat}}$ to $10^{-2}$ which is appropriate for MRI with net magnetic flux, we find that the midplane temperature reaches at most $2100\,$K. We also find that the saturated magnetic transport region (i.e. where the diffusivities are below the coupling threshold) only reaches out to $6\,R_J$, whereas the gravoturbulent region extends in as far as $120\,R_J$. However, we also note that increasing $\alpha_{\text{sat}}$ requires a further reduction of the  minimum value of  $\alpha$ to $2.2\times10^{-8}$ (at the boundary of the marginally coupled and gravoturbulent regions).

Opacity as a function of temperature is shown in the bottom--right panel of Fig. \ref{fig:models}, using the corresponding density profile [i.e., $\kappa(\rho(r), T(r))$ vs $T(r)$]. The opacity is complex and  varies by four orders of magnitude throughout the disk. Despite differences in the temperature and density profiles, the opacity profile for the self-consistent accretion disks follow  that of the constant-$\alpha$ disk. This is because the disks only deviate in the Grains opacity regime where the opacity is density independent (i.e., $a=0$).

\subsection{Ionisation}

\begin{figure*}
\begin{center}
\begin{tabular}{l r}
\includegraphics[width=0.46\textwidth]{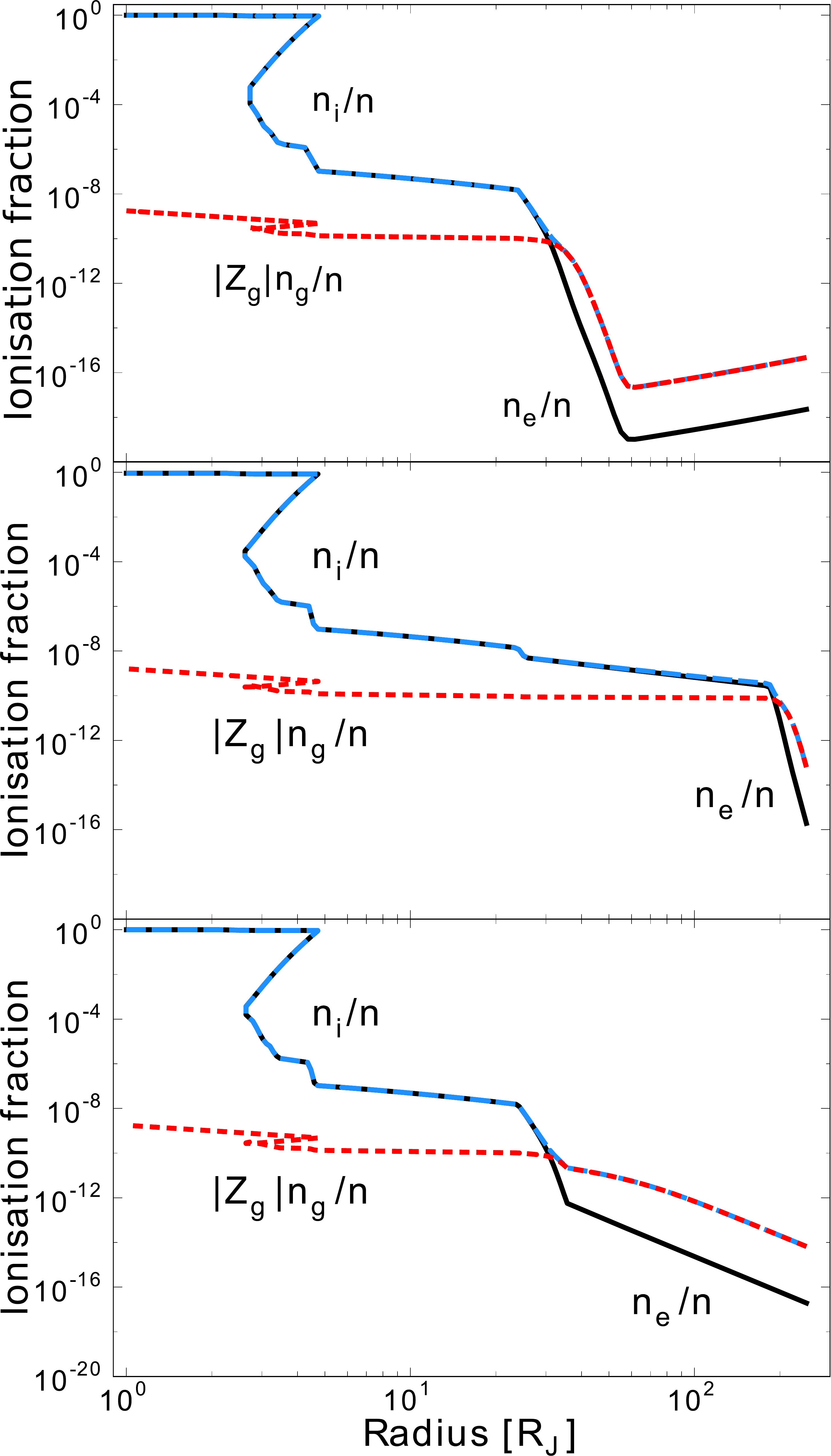} &\includegraphics[width=0.46\textwidth]{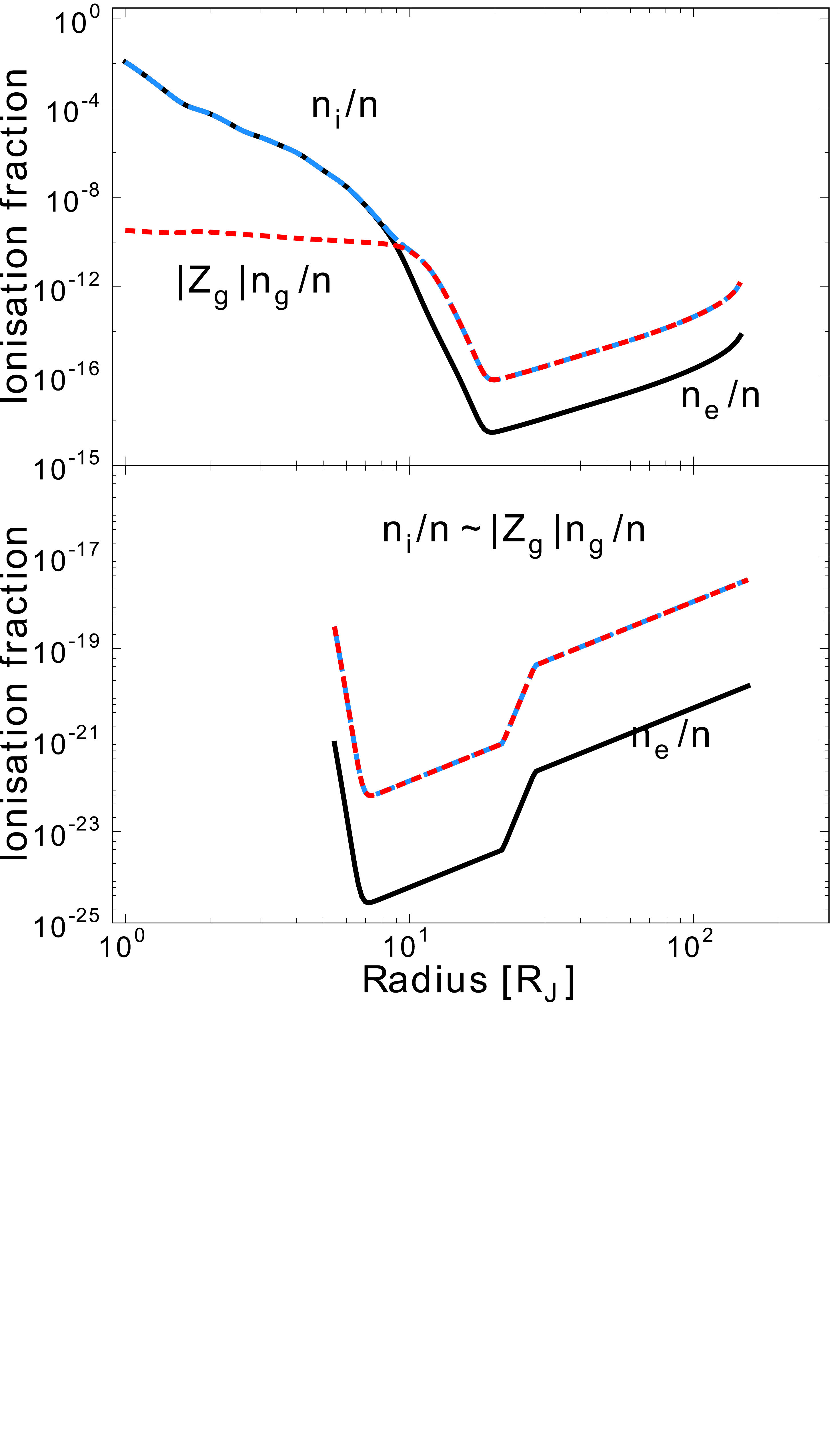}
\end{tabular}
\end{center}
\caption{Radial dependence of the midplane ionisation fraction, $n_e/n$ (solid curve), ion number density fraction, $n_i/n$ (dashed curve), and charge-weighted grain number density fraction, $|Z_g| n_g/n$ (dotted curve) for the constant-$\alpha$ model (top-left panel) and self-consistent accretion disk with MRI field (centre-left panel), and vertical field (bottom-left panel). For comparison, charged number density fractions are also shown for the Canup \& Ward $\alpha$ disk and MMJN in the top-right and bottom-right panels.} 
\label{fig:ionisation_fraction}
\end{figure*}

Fig. \ref{fig:ionisation_fraction} shows the electron (solid curve), ion (dashed curve), and charge-weighted grain (dotted curve) number density fraction for the constant-$\alpha$ model (top-left panel), and self-consistent accretion disks with MRI field (centre-left panel) and vertical field (bottom-left panel).  

In the constant-$\alpha$ disk, the ionisation fraction is high within the inner disk. Close to the planet the disk is almost fully ionised by thermal ionisation of hydrogen and helium, and thermal ionisation continues out to $\sim30\,R_J$ where the temperature exceeds $\sim1000$\,K and potassium is thermally ionised. In the abundance of free electrons grains acquire a large negative charge, $Z_g\sim-660$, but with little effect on the total electron density. Beyond this distance, the disk is not hot enough for significant thermal ionisation and so the ionisation fraction drops sharply. Ionisation is primarily by radioactive decay beyond $60\,R_J$, and the ionisation fraction is low (i.e., $n_e/n\sim10^{-19}$). In these conditions grains are mostly neutral, but still remove a large proportion of free electrons,  reducing the electron density by a factor of $\sim190$ relative to the ions. 

Thermal ionisation is strong over a larger portion of the self-consistent accretion disks, as the disk structure is reliant on a higher level of ionisation in the marginally magnetically coupled region.  We rely on thermal ionisation to achieve magnetic coupling, as midplane ionisation from radioactive decay, cosmic rays and X-rays is too weak (see \S\ref{sec:results_coupling}).  Grain charging is important beyond $\sim40\,R_J$ for both field geometries, however it has a greater effect for the vertical field where the ionisation fraction is lower. All profiles are multivalued between $3\,R_J\le r\le5\,R_J$, in keeping with the temperature profiles. 

Depletion onto grains removes heavy elements from the gas phase, and consequently reduces the ionisation fraction between $3\,R_J\lesssim r \lesssim 60\,R_J$ in the constant-$\alpha$ disk. There is no depletion close to the planet where ionisation is from the non-depleted hydrogen and helium, and in the outer disk  ionisation by radioactive decay is so weak that neutral metals are abundant  (i.e., $n_i/n_n\ll x_{\text{metals}})$ and the reaction rate is not limited by depletion.  In the intermediate region depletion reduces the ionisation fraction by up to the depletion factor, $10^{-\delta}=0.12$. The lowered electron density leads to a slight increase (up to $10\%$) in grain charge.  Depletion at this level has no appreciable effect on the structure of the self-consistent accretion disks.

Additional ionisation from MRI is ineffective for both the constant-$\alpha$ and fixed-temperature disks. Grain capture through vertical mixing rapidly removes ionisation in eddies produced in MRI active surface layers. If grains are absent, charges are removed by recombination quickly over a time-scale $\tau_R\approx4\Omega^{-1}$ at the outer edge.  However, if grains are present, even at the level $f_{dg}\gtrsim10^{-11}$, grain charge capture is rapid. Thus, free charges are rapidly removed as they are mixed into the dead zone and so do not contribute to midplane ionisation.

For ionisation produced through acceleration by MRI electric fields, we find that the electron energy is at most $\epsilon\approx5\times10^{-3}$\,eV in the constant-$\alpha$ disk, and lower still in higher density self-consistent accretion disks. This energy is orders of magnitude too low to ionise any atomic species.  Thus, there is no appreciable contribution from self-sustaining MRI ionisation in circumplanetary disks. Self-sustaining ionisation is more successful in protoplanetary disks where the density is lower such that electrons are able to be accelerated over a longer mean free path. 

We have also calculated the charge number density fractions for the Canup \& Ward $\alpha$ disk (top-right panel) and the MMJN (bottom-right panel) using the same method as given in \S\ref{sec:thermal_ionisation}. In the Canup \& Ward $\alpha$ disk thermal ionisation is high close to the planet with cosmic ray ionisation dominant beyond $20\,R_J$, similar to the constant-$\alpha$ disk. In the MMJN the ionisation fraction is very low ($n_e/n<10^{-16}$) due to both high surface density and low temperature.

\subsection{Magnetic field strength}

\begin{figure}
\begin{center}
\includegraphics[width=0.46\textwidth]{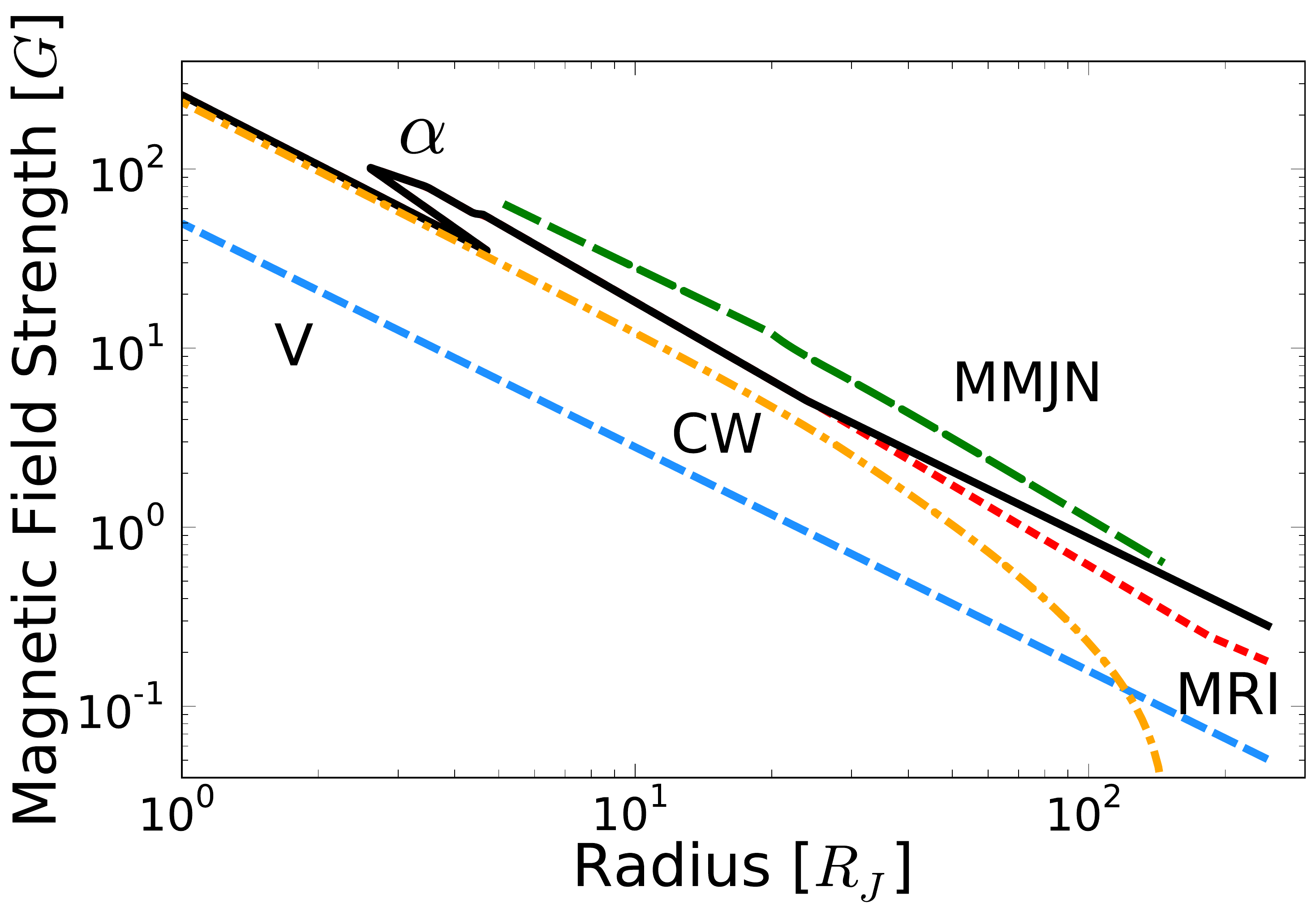} 
\end{center}
\caption{Radial dependence of the magnetic field strength, B, for the $\alpha$ model (solid curve), and fixed temperature model with MRI field (dotted curve), and vertical field (short-dashed curve), Canup \& Ward $\alpha$ disk (dot-dashed curve), and MMJN (long-dashed curve).}
\label{fig:magnetic}
\end{figure}

Fig. \ref{fig:magnetic} shows the magnetic field strength for the constant-$\alpha$ model (solid curve), and self-consistent accretion disks with MRI field (dotted curve) and vertical field (dashed curve).  

The MRI field strength for the constant-$\alpha$ disk varies between $B=0.28$--$250\,$G, and follows $B\propto r^{-1.1}$ across most of the disk. The field strength for the self-consistent  accretion disk with MRI field is almost identical to that of the constant-$\alpha$ disk, except for a small deviation at the outer edge where the temperature profiles diverge. The vertical field required for self-consistent  accretion has a similar dependency, with $B\propto r^{-5/4}$,  but it is $\sim$5 times weaker and decreases monotonically.  
All disk model fields are sub-equipartition and are consistent with the with the estimate of $B=10$--$50$\,G at $10R_J$ by \citet{2003A&A...411..623F}. 

We have plotted the magnetic field strength required to drive accretion throughout the entire disk for the self-consistent accretion disk with MRI field, however beyond $200\,R_J$ accretion is powered by gravitoturbulence rather than magnetic fields. We have no information about the magnetic field in the gravoturbulent region. 

For comparison we have calculated the MRI magnetic field strength for the Canup \& Ward $\alpha$ disk and the MMJN, which we also show in Fig. \ref{fig:magnetic}. We calculate the field strength the Canup \& Ward disk using equation (\ref{eq:alpha_magnetic_field}) for their $\alpha=6.5\times0^{-3}$, and for the MMJN using equation (\ref{eq:BMRI}) assuming an accretion rate of $\dot{M}=10^{-6}\,M_J/$year. 

\subsection{Magnetic coupling}
\label{sec:results_coupling}

\begin{figure*}
\begin{center}
\begin{tabular}{l r}
\includegraphics[width=0.46\textwidth]{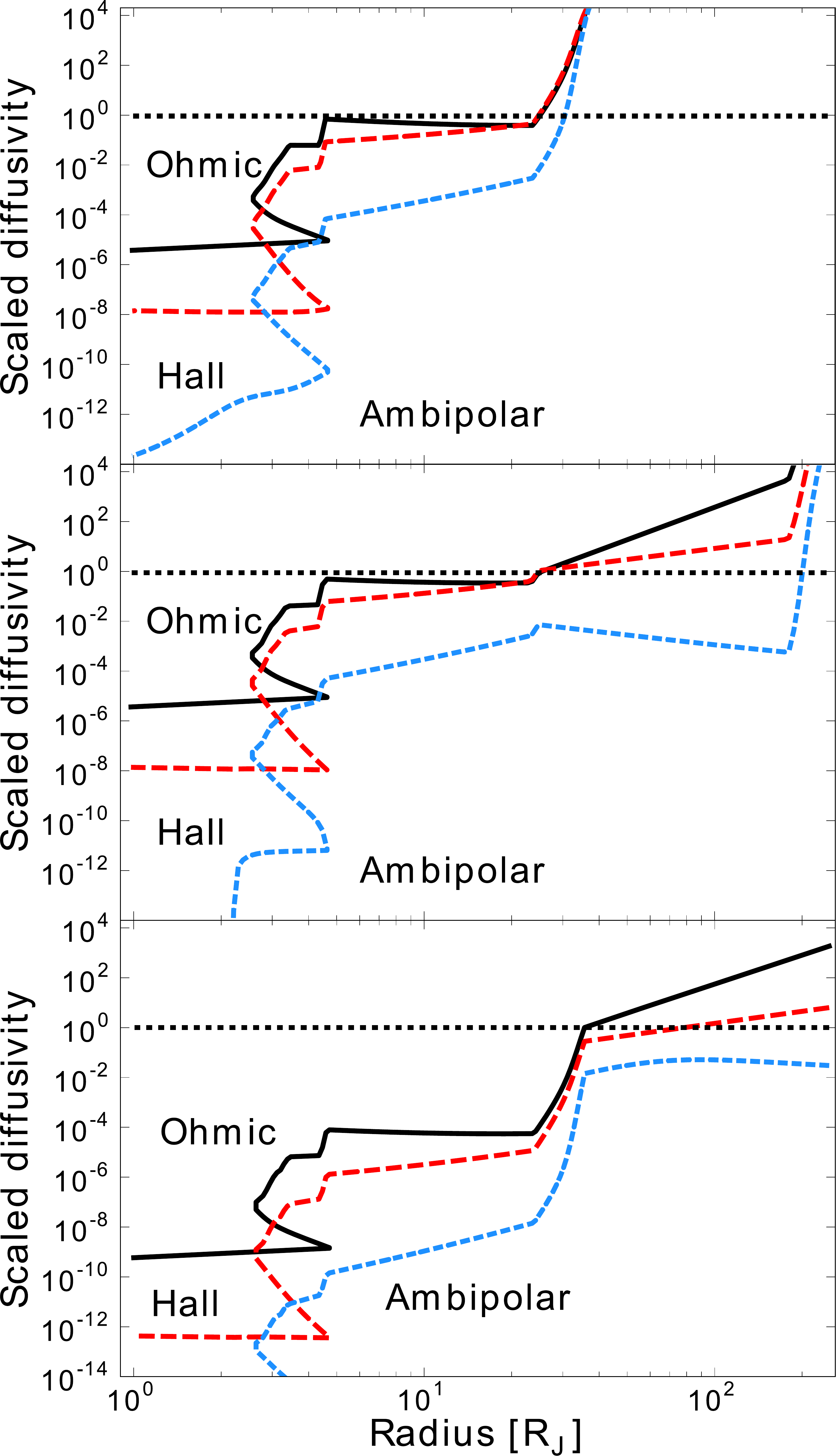}  & \includegraphics[width=0.46\textwidth]{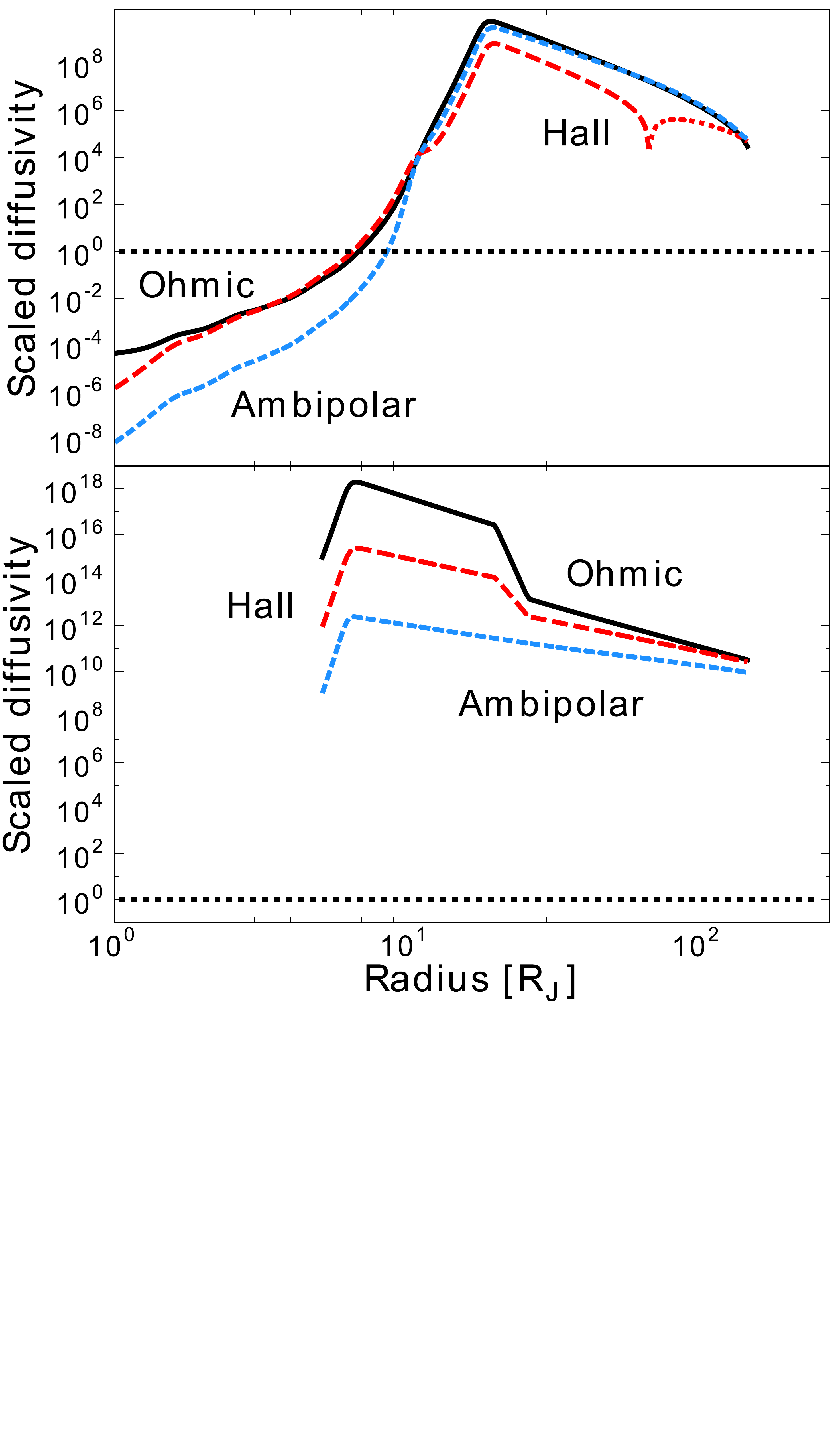}  
\end{tabular}
\end{center}
\caption{Radial  dependence of the Ohmic, $\eta_O$ (solid curve), Hall, $\eta_H$ (dashed curve) and Ambipolar, $\eta_A$ (dotted curve) diffusivities scaled by the coupling threshold (dotted horizontal line). The MRI field coupling threshold, $\eta/\Omega v_a^{-2}=1$,  is used for for the constant-$\alpha$ disk (top-left panel) and self-consistent accretion disk with MRI field (centre-left panel), while $\eta\Omega c_s^{-2}=1$, is used for the self-consistent accretion disk with vertical field (bottom-left panel). Diffusivities for the Canup \& Ward $\alpha$ disk and MMJN  are calculated for an MRI field in the top-right and bottom-right panels for comparison.}
\label{fig:diffusivity}
\end{figure*}

Fig. \ref{fig:diffusivity} shows the Ohmic (solid curve), Hall (dashed curve), and Ambipolar (dotted curve) magnetic diffusivities scaled by the coupling threshold for the constant-$\alpha$ disk (top panel), and self-consistent accretion disk with MRI field (centre panel) and vertical field (bottom panel).  The coupling threshold $\eta\Omega/v_a^2=1$ is used for the constant-$\alpha$ disk and self-consistent accretion disk with MRI field whereas $\eta\Omega/c_s^2=1$ is used for the self-consistent accretion disk with vertical field. The threshold is shown as a dotted horizontal line, with strong magnetic coupling in regions where each of the Ohmic, Hall and Ambipolar diffusivities are below the coupling threshold. 

We find that all disks are dense enough that Ohmic diffusivity dominates over Hall and Ambipolar. The diffusivities follow the inverse of the ionisation fraction [i.e., $\eta\propto n/n_e$, see equations (\ref{eq:ohmic})--(\ref{eq:ambipolar})]. Within $30\,R_J$, the ionisation fraction is high and so the diffusivities are well below the coupling threshold, $\eta\Omega v_a^{-2}\ll1$ or $\eta\Omega c_s^{-2}\ll1$ . At  $30\,R_J$  the diffusivities rise exponentially as thermal ionisation of potassium is suppressed by the low temperature. In the constant-$\alpha$ disk, ionisation from cosmic rays, X-rays and decaying radionuclides is too low for good magnetic coupling and so the majority of the disk, (i.e., $r>30\,R_J$), is uncoupled from the magnetic field.   The magnetically coupled region is larger at higher inflow rates where the midplane temperature is higher (i.e., the disk is coupled  within $90\,R_J$ for $\dot{M}=10^{-5}\,M_J/$year), however this also produces a higher disk scale height, (aspect ratio up to 0.79), violating the `thin-disk' approximation. Diffusivity below the coupling threshold in the inner disk indicates that the evolution of the disk and magnetic field are locked together, however the bulk of the disk is uncoupled to the magnetic field and accretion cannot proceed in these regions. 

The boundary of the magnetically coupled region is controlled by the exponential rise in the diffusivity at the ionisation temperature of potassium. For instance, if a vertical field is used instead of an MRI field, the scaled diffusivity is reduced by a factor $\left(v_a/c_s\right)^2=4\alpha$ [using the MRI field to evaluate $v_a$; see equation (\ref{eq:equipartition_ratio})], but the steepness of the diffusivity profile at the coupling boundary means that there is no change in the magnetically-coupled boundary. Similarly, depletion of heavy elements onto grains increases the diffusivity between $3\,R_J\le r \le 60\,R_J$, but does not change the radius of the magnetically-coupled region. 

The diffusivity profile for the self-consistent accretion disk with MRI field follows the constant-$\alpha$ disk profiles out until $30\,R_J$, where Ohmic diffusivity reaches the coupling threshold. Here, the disk enters the marginally magnetic coupled region and the rise in the diffusivity is not as steep. Although magnetic coupling is only weak, as the diffusivities are above the coupling threshold, it is still enough to drive accretion at the level given by equation (\ref{eq:alphaSS02}). This state of marginal coupling occurs out to $200\,R_J$, with Ohmic diffusivity up to $\sim10^4$ times greater than the coupling threshold. At the point where $Q=1$ gravitoturbulence becomes the dominant transport mechanism and the diffusivities resume their exponential rise. 

The coupling criterion for a vertical field is less stringent, and so the diffusivities are lower relative to the coupling threshold within $r\sim30\,R_J$.  As with the  self-consistent accretion disk with MRI field, the sharp rise in the diffusivity is reduced once the diffusivities reach the coupling threshold as the disk transitions to marginal magnetic coupling. However, in contrast, the disk never reaches $Q=1$ and so there is no transition to the gravoturbulent region.  

We have also calculated diffusivities for the Canup \& Ward $\alpha$ disk (top-right panel) and the MMJN (bottom-right panel) for an MRI field. We show the absolute value of the Hall diffusivity for the Canup \& Ward $\alpha$ disk as Hall diffusivity is negative beyond $r\sim70\,R_J$ (shown by a dotted curve when $\eta_H<0$). This occurs near the transition for ion re-coupling, and indicates that the Hall drift, between the field and neutrals, is in the opposite direction for a given field configuration. The diffusivities are above the coupling threshold for $r>10\,R_J$ for the Canup \& Ward $\alpha$ disk and at all radii for the MMJN, preventing magnetically-driven accretion in these regions.

\section{Discussion}
\label{sec:discussion}

In this paper we modelled steady-state accretion within a giant planet circumplanetary disk, and determined the effectiveness of magnetic fields and gravitoturbulence in driving accretion. We modelled the disk as a thin Shakura-Sunyaev $\alpha$ disk, heated by viscous transport and solved for the opacity simultaneously with the disk midplane structure using the \citet{2009ApJ...694.1045Z} opacity law, including the effects of self-gravity. Thermal ionisation dominates within $r\lesssim30\,R_J$ where the disk reaches the ionisation temperature of potassium ($T\sim10^3\,$K), but drops rapidly in  cooler  regions where ionisation is primarily by  radioactive decay. The midplane is too dense for penetration of cosmic rays or stellar X-rays. We considered both an MRI field and a vertical field in driving accretion, and found that a field of order $10^{-2}$--$10\,$G is needed to account for the inferred accretion rate onto the young Jupiter. To quantify the strength of interaction between the magnetic field and disk we calculated Ohmic, Hall, and Ambipolar diffusivities which cause slippage of the field lines relative to the bulk motion of the disk, decoupling their evolution. 

In the standard constant-$\alpha$ disk,  diffusivity is low enough for magnetic coupling in the inner region where potassium is thermally ionised.  However, the remainder of the disk is too cool for thermal ionisation and so strong diffusivity prohibits magnetically-driven accretion throughout the bulk of the disk. The disk is gravitationally stable, with Toomre's $Q\gg1$, and so there is no transport from gravitoturbulence either. 

This is inconsistent with the assumption of a constant-$\alpha$, and so we presented an alternate model in which $\alpha$ varies radially, ensuring that the accretion rate (taken to be uniform through the disk) is consistent with the level of magnetic coupling and gravitational instability. We achieved this by dividing the disk into three regions according to the mode of accretion: (i) the inner disk is hot enough for strong magnetic coupling through thermal ionisation and inflow is magnetically driven with $\alpha$ saturated at its maximum value; (ii)  Beyond $30\,R_J$ the disk is too cool for sufficient thermal ionisation of potassium and diffusivity exceeds the coupling threshold. Accretion is still magnetically driven, however as the magnetic coupling is weak, it occurs at a reduced efficiency with $\alpha$ inversely proportional to the level of magnetic coupling \citep{2002ApJ...577..534S}; (iii) The disk is gravitationally unstable in the outer regions where $Q\sim1$, and so gravitoturbulence is produced and drives accretion. Accretion is self-regulated so that the disk maintains marginal stability with $Q=1$. We calculated the disk structure for accretion driven by either MRI or vertical fields, finding very similar disk structures. With $Q\sim1$ at the outer edge, the disks are massive with $M_{\text{disk}}=0.5\,M_J$.

MHD analysis by \citet{2011ApJ...743...53F} and \citet{2013arXiv1306.2276T}  argue against magnetically driven accretion through the midplane where  the cosmic-ray and X-ray fluxes are too low. However, we find that midplane magnetic coupling relies primarily on thermal ionisation and so the disk temperature is crucial. \citet{2011ApJ...743...53F} use the surface temperature which is necessarily cooler than the midplane temperature, and so no thermal ionisation is expected.  \citet{2013arXiv1306.2276T} considers both MMJN models and actively supplied accretion disk models, appropriate for a later, and so cooler, phase than we consider here. MMJN models are necessarily cold to match conditions recorded by the final, surviving generation of Jovian moons, however these are likely formed late after a succession of earlier generations were accreted by the planet \citep{2006Natur.441..834C}. Temperatures in actively accreting disks are controlled by the inflow rate which likely decreases as inflow from the protoplanetary disk tapers. \citet{2013arXiv1306.2276T} consider inflow rates that are lower than ours by a factor of 5--70, so these disks model a cooler stage and consequently thermal ionisation is limited to the inner $4\,R_J$ of their highest inflow disk. Additionally, we also consider accretion in regions which are only marginally coupled to the magnetic field. We find that while saturated magnetic transport (i.e. with strong magnetic coupling) is limited to the inner $30\,R_J$, magnetically driven accretion with marginal coupling can potentially occur across the entire disk.

We have modelled steady-state accretion within the disk, with the assumption that the disk evolves toward or through this state during the proto-planet accretion phase. Numerical simulations indicate that accretion disks, including circumplanetary disks, rapidly evolve away from a self-gravitating state toward a quasi-steady state  \citep{2011MNRAS.410..994F, 2013ApJ...767...63S}, however there may be other time-dependent processes, such as short time-scale variability of inflow from the protoplanetary disk. Observations of accretion onto giant planets are needed to determine the accretion timescales, and how rapidly the accretion rate can change. 

The temperature profiles are multivalued in some regions of the disk, making the disks susceptible to viscous-thermal instability. This may lead to outbursts, undermining our steady-state assumption. This feature is only present when the inflow rate exceeds $\dot{M}=2\times10^{-8}\,M_J/$year, and so outbursting from the viscous-thermal instability will not occur at later time when the inflow rate has tapered off to below this value. While there is certainly the potential for outbursting at earlier times, our analysis centres on whether inflow driven by magnetic fields is plausible, rather than advocating a steady state solution.

There may also be additional torques on the circumplanetary disk, from stellar forcing or spiral waves generated by satellitesimals \citep{2012A&A...548A.116R, 2001ApJ...552..793G}, which we have not included. It is not clear what level of transport these processes produce during this phase of giant planet accretion and whether they can be incorporated as additional sources within the Shakura-Sunyaev $\alpha$ formalism.  We can model minor variations on the inflow parameters, such as a reduction in the accretion rate which reproduces the necessary cooling and disk mass lowering as inflow from the protoplanetary disk tapers. However these results are uncertain as they require yet lower values of $\alpha$ in the self-consistent accretion disk which are likely overwhelmed by the additional torques mentioned above.

Strong magnetic coupling near the surface of the planet will affect accretion onto the planet surface. The planetary magnetic field may channel the accretion flow onto the planet surface \citep{2011AJ....141...51L}, effecting the spin evolution of the planet \citep{1996Icar..123..404T, 2011AJ....141...51L}, and temperature of the planet.  However magnetospheric accretion requires diffusivity in order for the inflow to transfer onto the planetary magnetic field from the disk field. Loading onto the planetary field lines is only expected to occur close to the surface, if at all (at $r\sim1$--$3\,R_J$; see \citealt{1998ApJ...508..707Q, 2003A&A...411..623F, 2011AJ....141...51L}). However we find the diffusivity is very low at this distance, making loading of the gas onto the proto-planetary field lines from the disk field difficult. Magnetospheric accretion would require an additional source of diffusivity, such as  electron momentum exchange with ion acoustic waves (e.g., see \citealp{2006JGRA..111.1205P}), however it is not known how strong this effect is.

Finally, the circumplanetary disk is the formation site for satellites. The composition of the present day satellite systems around Jupiter and Saturn record conditions in their circumplanetary disks at the time of their formation. In particular, the rock/ice compositional gradients through the satellite systems set the disk ice line ($T\approx250\,$K)  at the the location of Ganymede, $r=15R_J$, and Rhea, $r=8.7\,R_S$, in the Jovian and Saturnian systems, respectively \citep{2003Icar..163..198M}. The location of the ice line is often incorporated or used as a measure of success in circumplanetary disk models (e.g., \citealt{1982Icar...52...14L, 2003Icar..163..198M, 2002AJ....124.3404C}), however no moons have been discovered beyond the Solar System and so it is not clear how typical the Jovian system is, nor to what degree these systems can vary \citep{2013arXiv1306.1530K}.  It is not our aim to reproduce the conditions for moon formation, but rather we are focussed on modelling the early phase of the disk, in which the disk is hot and there is the significant inflow onto Jupiter. Consequently, the ice line in our constant-$\alpha$ disk is at $r=139\,R_J$,  and the self-consistent accretion disks are too hot for ice. Several generations of satellites may have formed in these conditions, but the present day satellites likely form at a later stage when the disk has cooled as inflow into the circumplanetary disk tapers with the dispersal of the protoplanetary disk \citep{1989oeps.book..723C, 2006Natur.441..834C, 2010ApJ...714.1052S}.  Our results support the two stage circumplanetary disk evolution proposed by \citet{1989oeps.book..723C} in which the disk is initially hot and turbulent, but evolves to the cool quiescent disk as recorded by the giant planet satellite systems. 

In summary, we have found that during the final gas accretion phase of a giant planet the circumplanetary disk is hot and steady-state accretion may be driven by a combination of magnetic fields and gravitoturbulence. Accretion maintains the disk at a high temperature so that there is thermal ionisation through most of the disk. 

\section*{Acknowledgements}
We thank Yuri I. Fujii and Philippa K. Browning for valuable discussion and comments on the manuscript. We thank the anonymous referee for helpful comments which improved this manuscript. This work was supported in part by the Australian Research Council grant DP120101792. S.K. further acknowledges the support of an Australian Postgraduate Award and funding from the Macquarie University Postgraduate Research Fund scheme. This research has made use of NASA's Astrophysical Data System.  

\bibliographystyle{mn2e}
\bibliography{references}

\bsp
\label{lastpage}

\end{document}